\documentclass[review]{elsarticle}

\usepackage{hyperref}
\usepackage{longtable}
\usepackage{amsmath,amsthm,amssymb,epsfig,bm,amsmath}
 \usepackage{graphicx}
\usepackage{float}
\usepackage{amsmath}
\usepackage{mhchem}
\usepackage{xparse}
\usepackage{adjustbox}
\usepackage{MnSymbol}
\usepackage{mathdots}
\usepackage{makecell}
\usepackage{listings}
\usepackage{footnote}
\usepackage[flushleft]{threeparttable}
\usepackage{multirow}
\usepackage{relsize}
\usepackage{lettrine}
\usepackage{courier}
\usepackage{color} 
\definecolor{mygreen}{RGB}{28,172,0} 
\definecolor{mylilas}{RGB}{170,55,241}
\usepackage{amsmath,amsthm,amssymb,amsfonts} 
\usepackage{graphicx} 
\usepackage[margin=1in]{geometry} 
\usepackage[yyyymmdd]{datetime}
\usepackage{algorithm}
\usepackage[noend]{algpseudocode}
\usepackage{lipsum}
\usepackage{setspace}
\usepackage{siunitx}
\usepackage{pdfpages}
\usepackage{tikz}
\usetikzlibrary{positioning}
\usepackage{booktabs,caption}
\usepackage[]{hyperref}
\usepackage{soul}
\usepackage{xcolor}
\usepackage{subcaption}

\hypersetup{
 colorlinks  = true, 
 urlcolor   = blue, 
 linkcolor  = black, 
 citecolor  = blue 
}
\usepackage[numbers]{natbib}
\usepackage{color}
\definecolor{codegreen}{rgb}{0,0.6,0}
\definecolor{codegray}{rgb}{0.5,0.5,0.5}
\definecolor{codepurple}{rgb}{0.58,0,0.82}
\definecolor{backcolour}{rgb}{0.95,0.95,0.92}
\lstdefinestyle{mystyle}{
  backgroundcolor=\color{backcolour},  
  commentstyle=\color{codegreen},
  keywordstyle=\color{magenta},
  numberstyle=\tiny\color{codegray},
  stringstyle=\color{codepurple},
  basicstyle=\footnotesize,
  breakatwhitespace=false,     
  breaklines=true,         
  captionpos=b,          
  keepspaces=true,         
  numbers=left,          
  numbersep=5pt,         
  showspaces=false,        
  showstringspaces=false,
  showtabs=false,         
  tabsize=2,
  escapeinside={<@}{@>},
}
 
\usepackage{booktabs} 
\usepackage{siunitx} 
\usepackage{comment}
\usepackage{mathtools} 
\usepackage{gensymb} 
\usepackage{mhchem} 
\usepackage{fixltx2e} 
\usepackage{bbm}

\usepackage{multirow}

\usepackage{fancyhdr} 
\theoremstyle{definition}

\theoremstyle{definition}

\theoremstyle{remark}

\usepackage{soul} 
\soulregister\cite7
\soulregister\ref7
\soulregister\pageref7
\usepackage{bm}

\usepackage{framed} 

\usepackage{multicol} 

\usepackage{nomencl} 
\usepackage{tikz}
\makenomenclature
\usepackage[resetlabels,labeled]{multibib}

\renewcommand*\nompreamble{\begin{multicols}{2}}
\renewcommand*\nompostamble{\end{multicols}}

\usepackage{amssymb}
\usepackage{pifont}
\definecolor{light-gray}{gray}{0.95}

\usepackage[left, pagewise]{lineno}

\journal{Elsevier}

\begin{document}


\begin{frontmatter}

\title{\large Nuclear Microreactor Control with Deep Reinforcement Learning}

\author{Leo Tunkle$^{a,*}$, Kamal Abdulraheem$^{a}$, Linyu Lin$^{b}$, Majdi I. Radaideh$^{a,*}$}

\cortext[mycorrespondingauthor]{Corresponding Author: L. Tunkle (tunkleo@umich.edu), M. I. Radaideh (radaideh@umich.edu)}

\address{$^{a}$Department of Nuclear Engineering and Radiological Science, University of Michigan, Ann Arbor, Michigan 48109}
\address{$^{b}$Nuclear Science and Technology Division, Idaho National Laboratory, Idaho Falls, Idaho 83415}

\begin{abstract}
\small

The economic feasibility of nuclear microreactors will depend on minimizing operating costs through advancements in autonomous control, especially when these microreactors are operating alongside other types of energy systems (e.g., renewable energy). This study explores the application of deep reinforcement learning (RL) for real-time drum control in microreactors, exploring performance in regard to load-following scenarios. By leveraging a point kinetics model with thermal and xenon feedback, we first establish a baseline using a single-output RL agent, then compare it against a traditional proportional–integral–derivative (PID) controller. This study demonstrates that RL controllers, including both single- and multi-agent RL (MARL) frameworks, can achieve similar or even superior load-following performance as traditional PID control across a range of load-following scenarios. In short transients, the RL agent was able to reduce the tracking error rate in comparison to PID. Over extended 300-minute load-following scenarios in which xenon feedback becomes a dominant factor, PID maintained better accuracy, but RL still remained within a 1\% error margin despite being trained only on short-duration scenarios. This highlights RL’s strong ability to generalize and extrapolate to longer, more complex transients, affording substantial reductions in training costs and reduced overfitting. Furthermore, when control was extended to multiple drums, MARL enabled independent drum control as well as maintained reactor symmetry constraints without sacrificing performance---an objective that standard single-agent RL could not learn. We also found that, as increasing levels of Gaussian noise were added to the power measurements, the RL controllers were able to maintain lower error rates than PID, and to do so with less control effort. These findings illustrate RL's potential for autonomous nuclear reactor control, laying the groundwork for future integration into high-fidelity simulations and experimental validation efforts.

\end{abstract}


\end{frontmatter}


\setstretch{1.3}

\section*{Highlights}
\begin{itemize}
    \item Reinforcement Learning (RL) for load-following control in nuclear microreactors. 
    \item Exploration of multi-agent RL's potential for maintaining physical control actions. 
    \item Analysis of RL control's robustness under signal noise and generalization to new scenarios.
    \item Benchmarking of RL against other control methods such as PID. 
\end{itemize}

\section{Introduction}
\label{sec:intro}
Nuclear microreactors offer a pathway to reducing the financial risks of nuclear plant construction by lowering capital costs and enabling mass manufacturing for economies of scale. These compact reactors are being developed for applications such as remote power generation, coal power replacement, disaster relief, and load-following in support of renewable energy sources. However, their smaller power output limits potential revenue, making operational efficiency critical. Autonomous control reduces costs by minimizing onsite staffing, enabling remote deployment, and improving reliability under variable grid conditions \cite{shirvan2023uo2}.

Previous work in the area of autonomous nuclear control has mostly focused on traditional approaches such as proportional-integral-derivative (PID) control and model predictive control (MPC), due to their well-understood stability properties and widespread industrial adoption \cite{abdulraheem2025holos}. Since high-fidelity reactor modeling is computationally intensive, these studies generally use either a point kinetics model---a simplified, time-dependent reactor model that neglects spatial variation in the neutron population---or a multi-point kinetics model, which couples together a low number of point modeled regions. One point kinetics study aimed at improving upon PID control used genetic algorithms to optimize a fuzzy-PID control strategy so as to capture reactor response non-linearities at different power levels \cite{liu2009fuzzy}. In industry, Canada Deuterium Uranium (CANDU) reactors feature a publicized reactor regulating system based on an overarching control strategy composed of many proportional controllers \cite{CANDUinstruction}. To simulate this system, researchers developed a multi-point kinetics model with 14 zones \cite{javid2008candu}. Machine learning (ML) for control purposes has been explored for decades, often within traditional frameworks. In the early 1990s, for example, researchers trained forward and diagonal recurrent neural networks by using linear MPC outputs to optimize temperature control in a point kinetics reactor model \cite{ku1992drnn}.

More recent work has looked into applying control strategies to drum-controlled reactors, which are of increasing relevance  thanks to the Westinghouse and HolosGen microreactor designs that feature control drums for regular operation---instead of only control rods, as is the case in most operating reactors. A study that used a point kinetics model of a drum-controlled 1 MWe space nuclear reactor applied fuzzy-logic-based control to improve robustness to external disturbances while also considering the physical constraints of drum motion \cite{zhao2023spacedrum}. Another study investigating a 300 MWt space reactor for nuclear thermal propulsion was unique in that it did not use point kinetics \cite{laboure2023ntp}. In comparing a hybrid PID controller that accounts for both power and reactivity in light of period-generated control \cite{bernard1992period}, the present work uses a much higher-fidelity MOOSE \cite{moose} model that couples full dimensional neutronics and thermal hydraulics models of the reactor. In regard to optimization of control drum configurations in the HOLOS-Quad microreactor design, moth-flame optimization was shown to be a good real-time control strategy in comparison to five other optimization algorithms that were also studied \cite{price2022moth}.

Outside the nuclear domain, while PID control and MPC continue to dominate in practical applications, reinforcement learning (RL) is starting to enable real-world improvements. Boston Dynamics' robot dog Spot, which primarily leverages MPC, was upgraded to utilize RL to improve its stability in hard-to-model scenarios such as walking on greasy surfaces \cite{BDblog}. RL control has been successfully applied to commercial cooling systems, achieving energy savings of 9\%--13\% in comparison to conventional controllers \cite{luo2022cooling}. In that particular case, MPC was specifically ruled out due to the impracticality of using physics models for large building systems. Furthermore, an offline model-based RL framework trained on historical operation data was successfully deployed in four coal power plants in order to increase combustion efficiency \cite{zhan2022coalplant}. MPC for real-time control was once again ruled out, this time due to the large scale of the control system in question. On smaller scales, studies have demonstrated RL control's effectiveness at handling uneven loading in washing machines, halving the undesirable vibrations and noise in heavily loaded machines \cite{shimizu2022washer, kang2024washer}. When applied to optimization more broadly, RL has successfully outperformed humans at computer chip floorplanning \cite{mirhoseini2021chips}, identifying matrix multiplication algorithms whose computational efficiencies exceed those of previously known algorithms \cite{fawzi2022matrix}, and reducing bitrate consumption for video buffering compared to heuristics and supervised learning \cite{gauci2019fbvideo}.

The efficacy of RL has not gone unnoticed in the nuclear field, with several papers having applied it to nuclear-relevant optimization and control problems. RL was shown to outperform a genetic algorithm in determining the optimal placement of detectors for monitoring the 3-D neutron flux distribution in a reactor core \cite{tan2024detector}. For the generally static problem of reactor design optimization, RL algorithms in what is now an open-source framework \cite{radaideh2023neorl} were used to optimize small-\cite{radaideh2021physics} and large-scale assembly designs \cite{radaideh2021large}, as well as full nuclear reactor cores \cite{seurin2024multi,seurin2024physics}. RL concepts have also played a role in enhancing the performance of metaheuristic algorithms through neuroevolutionary algorithms \cite{radaideh2021rule} and experience replay techniques \cite{radaideh2022pesa}. These studies all highlight the efficiency of RL in exploring large combinatorial spaces.

In control settings, the experience-hungry nature of RL necessitates the use of low-fidelity reactor models to keep training costs reasonable. For reactivity control via rods in a boiling-water reactor, the deep deterministic policy gradient algorithm was shown to outperform $H_\infty$ control under external disturbances \cite{chen2022deep}. That study modeled the reactor by using point kinetics with one delayed neutron group and fuel temperature reactivity feedback. The fact that there was no modeling of xenon poisoning was justifiable due to the short timescales examined. Using a fast-running simulator from the Korea Atomic Energy Research Institute \cite{Kwon1997CNS}, a multi-output control policy for four valves and a heater was trained using the soft actor-critic algorithm in combination with hindsight experience replay to successfully meet important control targets and safety requirements during the heat-up mode of a nodally modeled pressurized-water reactor \cite{bae2023multiPWR}. A study comparing two RL algorithms in terms of finding optimal microreactor control drum positions for power symmetry and criticality in a time-varying fuel burnup scenario necessitated the use of a surrogate model for fidelity, due to the multi-year timescales involved \cite{radaideh2025multistep}. This surrogate model was itself a deep neural network trained on thousands of datapoints generated through intensive Monte Carlo core simulations.

None of the previously covered literature included a study demonstrating RL algorithms' potential for real-time load-following control in nuclear microreactors, which feature control mechanisms different from traditional GW-scale nuclear power plants. The present paper advances RL for autonomous nuclear reactor control in three key ways. First, we demonstrate feasibility via the first application of RL to real-time drum-based reactor control, and we systematically evaluate the performance of single-agent RL across a variety of test scenarios, in comparison to traditional PID control. Second, we introduce a multi-agent RL (MARL) framework that leverages reactor symmetry to enhance training efficiency, improve controller performance, and unlock the full potential of drum control. By demonstrating MARL’s generalizability and its satisfaction of physical constraints, this work lays the foundation for RL-based control in more complex reactor systems and operating conditions. Third, we evaluate the resilience of the trained RL control agents---both individually and in multi-agent configurations---against noise and uncertainties in the input signals so as to assess their effectiveness. \textit{Overall, this study represents a pioneering investigation into the potential and limitations of RL for controlling drum-based nuclear microreactor systems}.

\section{Description of the Microreactor Model}
\label{sec:reactor}
\subsection{Holos-Quad Microreactor}
\label{sec:reactordesign}
The Holos-Quad microreactor \cite{lee2020holoscore}\cite{anton2022holos}, developed by HolosGen LLC, is a high-temperature gas-cooled reactor designed for scalable, self-contained power generation. Its architecture was inspired by closed-loop turbojet engines, replacing conventional combustion chambers with sealed nuclear fuel cartridges. This integrates fuel, moderation, heat exchange, and power conversion within individual pressure vessels, termed ``subcritical power modules'' (SPMs), thus eliminating the need for external coolant loops or a traditional balance of plant. The entire system is compact enough to be housed within a 40-ft ISO container, making it transportable via standard commercial methods, including by truck or airlift. The reactor is intended for use in remote locations, or in disaster relief scenarios where rapid deployment and autonomous operation are desired.

The Holos-Quad system uses four SPMs, each arranged symmetrically around a central reflector. Each SPM remains subcritical when isolated and becomes critical only when coupled close together with a positioning system, thereby ensuring a degree of passive safety. Additionally, shutdown rods in the central reflector provide active safety \cite{kinast2024parametric}. High-assay low-enriched uranium (HALEU [19.95\% U-235]) fuel is encapsulated within tri-structural isotropic (TRISO) particles, which are themselves contained within fuel channels in the graphite moderator blocks, providing layers of resistance against high temperatures and fission product release. Fuel is loaded during fabrication of the SPM, enabling 3–20 years of continuous operation before refueling or disposal, depending on operating conditions. The helium coolant circulates at high pressures, transferring heat through a separate Brayton cycle within each SPM. In addition to the Brayton cycle, a secondary organic Rankine cycle heat recovery system increases the overall thermal efficiency to 45\%–60\%.

\begin{figure}[!h]   
\centering
\includegraphics[width=0.9\textwidth]{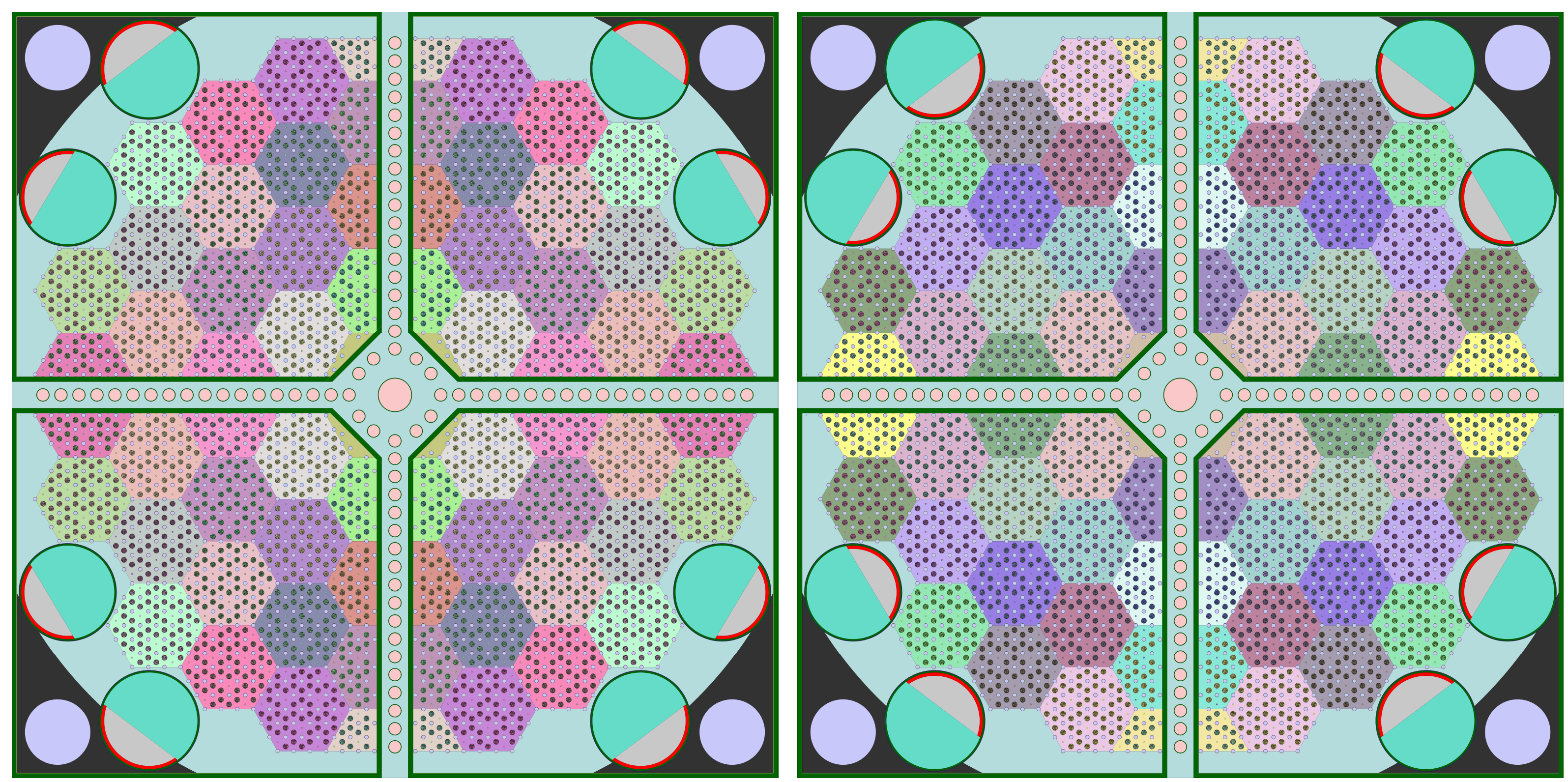}\\[-0.2cm]
\caption{Axial slice of the Holos-Quad design with control drums out (left) and control drums inserted (right).}
\label{fig:holos}
\end{figure}

Reactivity control is achieved through eight cylindrical control drums that are embedded and actuated in the outer regions of the core and that each contain a boron carbide (B$_4$C) neutron absorbing layer. These drums regulate reactivity insertion to control the neutron population and therefore the reactor power, thus allowing for load-following capabilities. Fig.~\ref{fig:holos} depicts a cross section of the reactor. The control drums are seen to be rotated fully outward, such that the neutron absorbing portion of the drum is as far from the core as possible; as well as fully inward, such that the maximum number of neutrons are absorbed to decrease the power.

\subsection{Reactor Modeling}
\label{sec:reactormodel}
Modeling the dynamic response of nuclear reactors is central to developing effective control strategies, particularly for applications such as load following, in which the power output must adapt to changing external demands. The complexity of reactor physics necessitates models that balance computational efficiency and fidelity in capturing key feedback mechanisms.

One of the most widely used approaches in reactor control research is the point kinetics model \cite{ott1985}, which simplifies the full spatial neutron transport problem by assuming a single effective neutron population for the entire core. Despite its limitations in spatial resolution, the point kinetics approach is computationally efficient and can accurately capture global reactor power dynamics under transient conditions. The standard six-group delayed neutron point kinetics equations describe the evolution of neutron density and precursor concentrations:

\begin{equation}
\frac{d\bar{n}(t)}{dt} = \frac{\rho(t) - \beta}{\Lambda} \bar{n}(t) + \sum_{i=1}^{6} \frac{\beta_i}{\Lambda} \bar{C}_i (t),
\label{eq:pke}
\end{equation}

\begin{equation}
\frac{d\bar{C}_i (t)}{dt} = \lambda_i \bar{n}(t) - \lambda_i \bar{C}_i (t), \quad i=1,2,\dots,6,
\end{equation}
where $\bar{n}(t)$ is the normalized neutron density, $\bar{C}_i (t)$ is the normalized concentration of the \textit{i}th precursor, $\rho(t)$ is the reactivity, $\beta$ is the total delayed neutron fraction, $\beta_i$ is the delayed neutron fraction for the \textit{i}th precursor, $\lambda_i$ is the decay constant of the \textit{i}th precursor, and $\Lambda$ is the neutron generation time.

In addition to the insertion of reactivity due to control mechanisms, thermal feedback effects and xenon poisoning must be included to capture transient reactor behaviors. These are accounted for in the following reactivity model:

\begin{equation}
\delta \rho(t) = \delta \rho_d(t) + \alpha_f \delta T_f(t) + \alpha_m \delta T_m (t) + \alpha_c \delta T_c (t) - \delta \rho_X (t),
\end{equation}
where $\delta \rho_d$ represents the reactivity insertion from control drums; $\alpha_f, \alpha_m,$ and $\alpha_c$ are the reactivity coefficients for the fuel, moderator, and coolant temperatures, respectively ($T_f$, $T_m$, and $T_c$); and $\delta \rho_X$ accounts for the xenon poisoning effects.

The nuclear power plants in operation today generally insert and remove neutron-absorbing control rods axially to control the reactor's neutron population. The Holos-Quad microreactor also has control rods available for emergency shutdown scenarios, but otherwise relies on control drums instead. In this reactor, the total reactivity insertion from control drums can be approximated using an analytic cosine-based formulation for each drum, as derived from first-order perturbation theory:

\begin{equation}
\rho_d(\theta) = \frac{\rho_{d,\max}}{2} (1 - \cos\theta),
\label{eq:drumworth}
\end{equation}
where $\rho_{d,\max}$ is the maximum inserted reactivity (corresponding to a $180^\circ$ rotation fully outward) and $\theta$ is the control drum angle. This approximation was previously formulated and verified by using the Serpent \cite{serpent} Monte Carlo code to simulate the Holos-Quad core---with various combinations of drum angles---and to determine the overall and differential control drum worths \cite{kochunas2019holosdrum}. Conducting reactivity analysis and control rod worth estimation using Monte Carlo codes is a well-established and state-of-the-art method, as demonstrated by numerous studies \cite{brunton2016sparse,radaideh2018reactivity,sembiring2021analysis,fejt2022utilization}. The corresponding differential control drum worth was thus similarly confirmed to take the following expression:

\begin{equation}
\frac{d\rho_d(\theta)}{d\theta} = \frac{\rho_{\max}}{2} \sin\theta.
\label{eq:diffdrumworth}
\end{equation}

Keeping with the reactor point approximation used for the neutron population, thermal feedback may be modeled using a three-temperature lumped heat balance formulation \cite{abdulraheem2025load}:

\begin{equation}
m_f c_f \frac{dT_f}{dt} = q P_r \bar{n}(t) - K_{fm} (T_f - T_m),
\end{equation}

\begin{equation}
m_m c_m \frac{dT_m}{dt} = (1-q) P_r \bar{n}(t) + K_{fm} (T_f - T_m) - K_{mc} (T_m - T_c),
\end{equation}

\begin{equation}
m_c c_c \frac{dT_c}{dt} = K_{mc} (T_m - T_c) - \dot{m}_c c_c (T_c - T_{\text{in}}),
\end{equation}
where $m_f, m_m, m_c$ and $c_f, c_m, c_c$ are the masses and heat capacities of the fuel, moderator, and coolant, respectively; $P_r$ is the reactor’s rated power; $q$ is the fraction of energy from fission products deposited in the fuel; $K_{fm}$ and $K_{mc}$ are heat transfer coefficients; $\dot{m}$ is the mass flow rate of the coolant; and $T_{\text{in}}$ is the inlet coolant temperature.

On load-following timescales of hours or more, xenon buildup and decay are important in accurately modeling core reactivity. Xenon-135, a strong neutron absorber, is produced through fission and iodine-135 decay, and is removed through beta decay and transmutation via neutron capture:

\begin{equation}
\frac{dI(t)}{dt} = \gamma_I \Sigma_f v \bar{n}(t) - \lambda_I I(t),
\end{equation}

\begin{equation}
\frac{dX(t)}{dt} = \gamma_X \Sigma_f v \bar{n}(t) - \lambda_X X(t) + \lambda_I I(t) - \sigma_X v \bar{n}(t) X(t).
\end{equation}
where $I(t)$ and $X(t)$ represent the iodine and xenon concentrations, respectively; $\gamma_I$ and $\gamma_X$ are their respective fission yields; $\Sigma_f$ describes the macroscopic fission cross section; $v$ is the average velocity of thermal neutrons; $\sigma_X$ is the microscopic absorption cross section of xenon; and $\lambda_I$ and $\lambda_X$ are decay constants. The xenon reactivity may then be modeled as:

\begin{equation}
\delta \rho_X (t) = -\sigma_X v X(t).
\label{eq:rhoX}
\end{equation}

Overall, the modeling methods described herein are limited by the use of point kinetics, which sacrifices spatial fidelity in return for simulation speed but still maintains the overall behavior of the neutron population by implementing control drum actions. Rapid simulation speed is necessary to supply RL algorithms (and controllers in general) with sufficient experience to learn good control policies. Table \ref{tab:Holosparameter} lists all the parameters relevant to this application of the Holos-Quad microreactor model.

\begin{table}[htbp]
    \centering
    \caption{HolosGen microreactor parameters used in the point kinetics model.}
    \begin{tabular}{|c|c|c|c|c|c|}
        \hline
        Parameter & Value & Unit & Parameter & Value & Unit \\
        \hline
        $\beta$ & 480.10 & pcm & $\alpha_f$ & -2.875 & pcm/K \\
        \hline
        $\beta_1$ & 14.20 & pcm & $\alpha_m$ & -3.696 & pcm/K \\
        \hline
        $\beta_2$ & 92.40 & pcm & $\alpha_c$ & 0.000 & pcm/K \\
        \hline
        $\beta_3$ & 78.00 & pcm & $c_{p,f}$ & 977.00 & J/Kg/K \\
        \hline
        $\beta_4$ & 206.60 & pcm & $c_{p,m}$ & 1697.00 & J/Kg/K \\
        \hline
        $\beta_5$ & 67.10 & pcm & $c_{p,c}$ & 5188.60 & J/Kg/K \\
        \hline
        $\beta_6$ & 21.8 & pcm & $m_f$ & 2002.00 & Kg \\
        \hline
        $\Lambda$ & 0.00168 & s & $m_m$ & 11573.00 & Kg \\
        \hline
        $\lambda_1$ & 0.01272 & 1/s & $m_c$ & 500.00 & Kg \\
        \hline
        $\lambda_2$ & 0.03174 & 1/s & $\dot{m}_c$ & 17.50 & Kg/s \\
        \hline
        $\lambda_3$ & 0.11600 & 1/s & $T_{f0}$ & 832.4 & K \\
        \hline
        $\lambda_4$ & 0.31100 & 1/s & $T_{m0}$ & 830.22 & K \\
        \hline
        $\lambda_5$ & 1.40000 & 1/s & $T_{in0}$ & 795 & K \\
        \hline
        $\lambda_6$ & 3.87000 & 1/s & $T_{out0}$ & 1106 & K \\
        \hline
        $n_0$ & 2.25E+13 & m$^{-3}$ & $K_{fm}$ & 1.17E+06 & W/K \\
        \hline
        $P_r$ & 22.00 & MW & $K_{mc}$ & 2.16E+05 & W/K \\
        \hline
        $q$ & 0.96 & - & $\sum_f$ & 0.1117 & - \\
        \hline
        $\gamma_I$ & 0.061 & - & $\gamma_x$ & 0.002 & - \\
        \hline
        $\lambda_I$ & $2.87 \times 10^{-5}$ & $s^{-1}$ & $\lambda_x$ & $2.09 \times 10^{-5}$ & $s^{-1}$ \\
        \hline
        $\upsilon$ & 2.19e+3 & m/s & $\sigma_x$ & $2.65 \times 10^{-22}$ & $cm^2$ \\
        \hline
        $\rho_{d,\max}$ & 511 & pcm & - & - & - \\
        \hline
    \end{tabular}
    \label{tab:Holosparameter}
\end{table}

\section{Background on Reinforcement Learning}
\subsection{Conventional Reinforcement Learning}
\label{sec:rlbackground}
\begin{figure}[!h]
\centering
\includegraphics[width=0.9\textwidth]{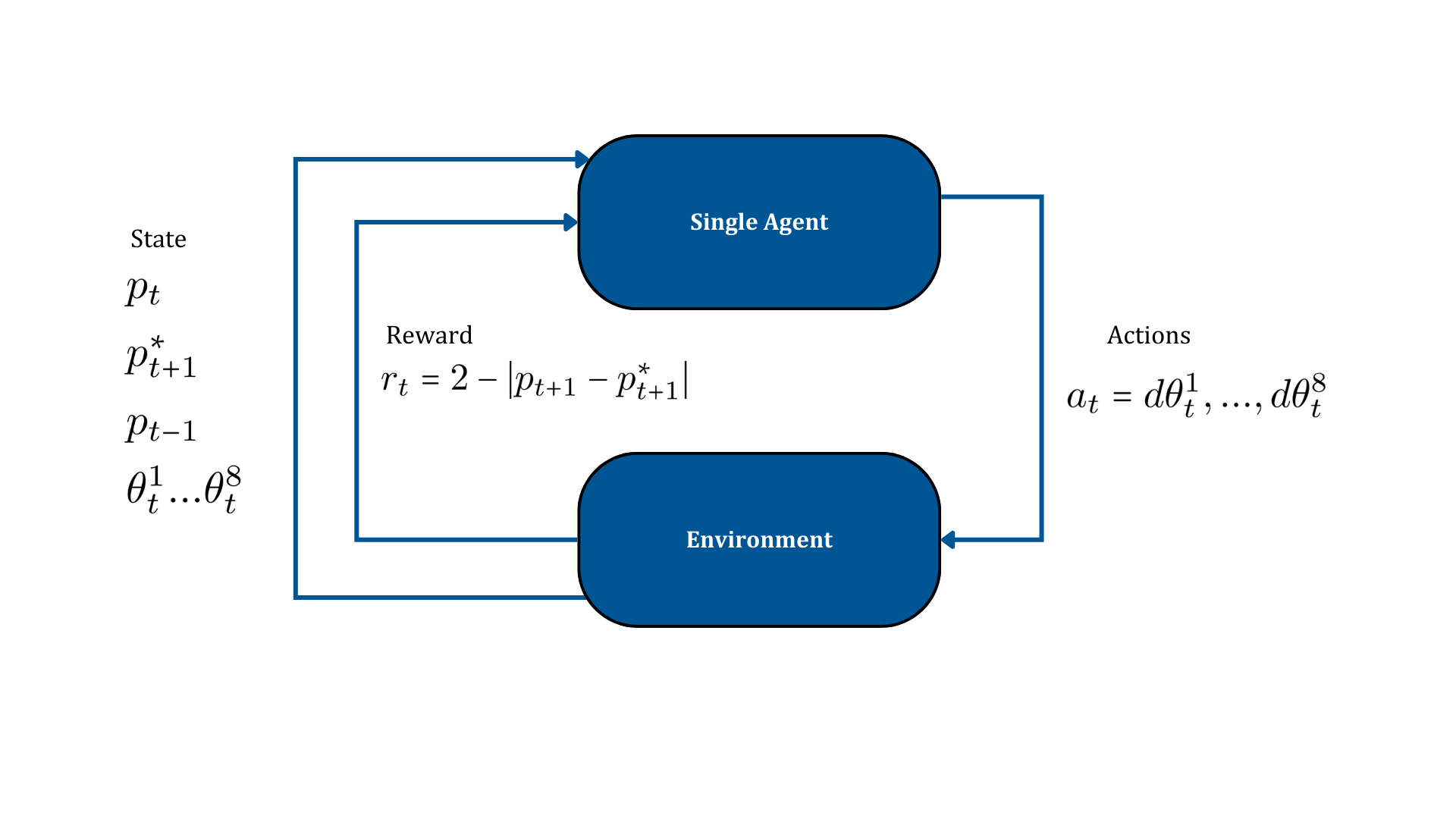}\\[-0.2cm]
\caption{RL training loop, where $P^*$ is the target reactor power, $P$ is the simulated reactor power, and $d\theta$ is the control drum rotation speed.}
\label{fig:rl_loop}
\end{figure}

RL is an ML approach that trains an agent to interact with an environment by performing actions that maximize a given long-term objective \cite{suttonbarto}. In control systems, the ``agent'' corresponds to the controller, and the ``environment'' represents the system being controlled. Fig.~\ref{fig:rl_loop} visualizes the training loop specific to this paper.

During training, the agent observes the system state at time $s_t$ and selects an action $a_t$. This action causes the environment to transition to a new state $s_{t+1}$, and the agent then receives feedback in the form of a reward $r_{t}$, as defined by a reward function. The agent repeats this process, resulting in a trajectory of states, actions, and rewards: $s_0, a_0, r_0, s_1, a_1, r_1, s_2, a_2, r_2$, etc. A trajectory that is finite as a result of problem scope or truncation is known as an episode.

Initially, the agent selects poor actions, but ideally these will converge over time, through trial and error, toward an optimal policy that chooses actions that maximize the cumulative episode reward. Exactly how this improvement occurs depends on the RL algorithm used. Modern RL algorithms primarily fall into the following three categories:
\begin{enumerate}
    \item Value-based methods such as Q-learning \cite{watkins1992qlearning} approximate the expected reward of each state-action pair and act greedily with respect to these estimates.
    \item Policy-based methods such as policy gradient methods \cite{sutton1999policygradient} directly optimize a parameterized policy to maximize the reward.
    \item Actor-critic methods such as Proximal Policy Optimization (PPO) \cite{schulman2017ppo} combine the strengths of both approaches by training a policy (actor) while simultaneously estimating state values (critic) to reduce variance in learning.
\end{enumerate}

Since implementation of an RL environment is application-specific, a common interface is required to allow RL algorithms to work generally across diverse applications. The \textit{Gymnasium} \cite{towers2024gymnasium} Python library has become the standard for implementing RL environments. In this framework, the environment provides an interface via which it receives actions submitted by the agent, computes the system state and reward that would result from those actions, and returns these back to the agent. This modular structure standardizes benchmarking and enables development of RL algorithms applicable to different environments, provided they each follow the \textit{Gymnasium} interface.

PPO in particular is known for stable learning and reduced hyperparameter sensitivity in comparison to older methods such as Deep Q-Networks \cite{mnih2013dqn} or Trust Region Policy Optimization \cite{schulman2015trpo}. PPO clips the policy gradient to prevent excessively large policy updates, thus enhancing learning stability. The \textit{Stable-Baselines3} \cite{sb3} library offers a widely adopted PPO implementation that emphasizes reliability and ease of use. Unlike self-implementations, \textit{Stable-Baselines3} provides well-tested algorithms, consistent interfaces, and built-in logging tools, making it a standard choice for research and practical applications. At a high level, its PPO implementation uses a single neural network for both actor and critic, and this network is updated during training to maximize the following objective function:
\begin{equation}
    L^{total}(\theta) = L^{policy}(\theta) - c_1 L^{value}(\theta) + c_2 H,
    \label{eq:ppo}
\end{equation}
where $\theta$ represents the network weights; $L^{policy}(\theta)$ represents how well the agent performed, based on the rewards received; $L^{value}(\theta)$ represents the accuracy with which the agent can judge the value of a given state, and is used to stabilize training; and $H$ is an entropy term to help the agent continue to explore the action space and not get tied too early to a non-optimal strategy. Thus, $c_1$ is a hyperparameter that balances how much weight is placed on the critic output versus the actor output of the neural network. Its default setting is 0.5 so as to weigh the two equally. Also by default, $c_2$, the entropy hyperparameter, is chosen to be 0. This is because, during training, actions are chosen stochastically based on a probability distribution, and this serves a similar purpose as the entropy term.

Many other hyperparameters exist, but since PPO has been shown to be relatively robust to hyperparameter tuning, we leave further details to other resources and only describe three others: $N_{envs}$, $N_{steps}$, and $N_{timesteps}$. Parallelization of training is made possible by instantiating a so-called ``vectorized'' environment, which runs $N_{envs}$ environments in parallel, thus providing a vector of observations to, and receiving a vector of actions from, the agent. In this way, the amount of experience from which the agent can learn in a given time frame is multiplied, speeding up the learning process. However, depending on computing resources and environment complexity, at a certain point the bottleneck becomes message-passing and agent network updates rather than accumulation of experience. $N_{steps}$ is the number of timesteps collected into a rollout buffer per environment before the buffer is used to update the agent network via the objective function. This means that policy updates occur after $N_{envs} \times N_{steps}$ timesteps of experience have been accumulated. $N_{timesteps}$ is the total number of timesteps to train for, over which $\frac{N_{timesteps}}{N_{envs} \times N_{steps}}$ policy updates will occur. This training loop is expressed in Algorithm Block \ref{alg:ppo}.

\begin{algorithm}[!h]
  \small
    \caption{Single-agent RL for microreactor control.}
    \label{alg:ppo}
    \begin{algorithmic}[1]
     \State \textbullet Initialize network and thus policy $\pi_{\theta_{old}}$ with random weights, $\theta_{old}$
     \State \textbullet Initialize $r^{best}$=$-\infty$
     \For{Rollout $i = 1$ to $\frac{N_{timesteps}}{N_{envs} \times N_{steps}}$} 
             \For{Env $j = 1$ to $N_{envs}$}
                \For{Timestep $t = 1$ to $N_{steps}$}
                    \State \textbullet Run policy $\pi_{\theta_{old}}$ and collect rewards
                \EndFor 
            \EndFor 
        \State \textbullet  Optimize objective $L^{PPO}(\theta)$, Eq.\eqref{eq:ppo}, with respect to $\theta$
        \State \textbullet Set $\theta_{old} = \theta$
        \State \textbullet Calculate mean reward $\bar{r}_i$ for the current rollout. See Section \ref{sec:interfacemetrics}.
        \If {$\bar{r}_i > r^{best}$}
            \State \textbullet  Save the current network parameters, $\theta$
            \State \textbullet  Set $r^{best} = \bar{r}_i$
        \EndIf
     \EndFor 
    \end{algorithmic}
\end{algorithm}

\subsection{Multi-Agent Reinforcement Learning}
\label{sec:marlbackground}
MARL extends RL to environments where multiple agents can learn and interact simultaneously. These interactions may be cooperative, competitive, or mixed, depending on the task structure and reward design.

\begin{figure}[!h]
\centering
\includegraphics[width=0.9\textwidth]{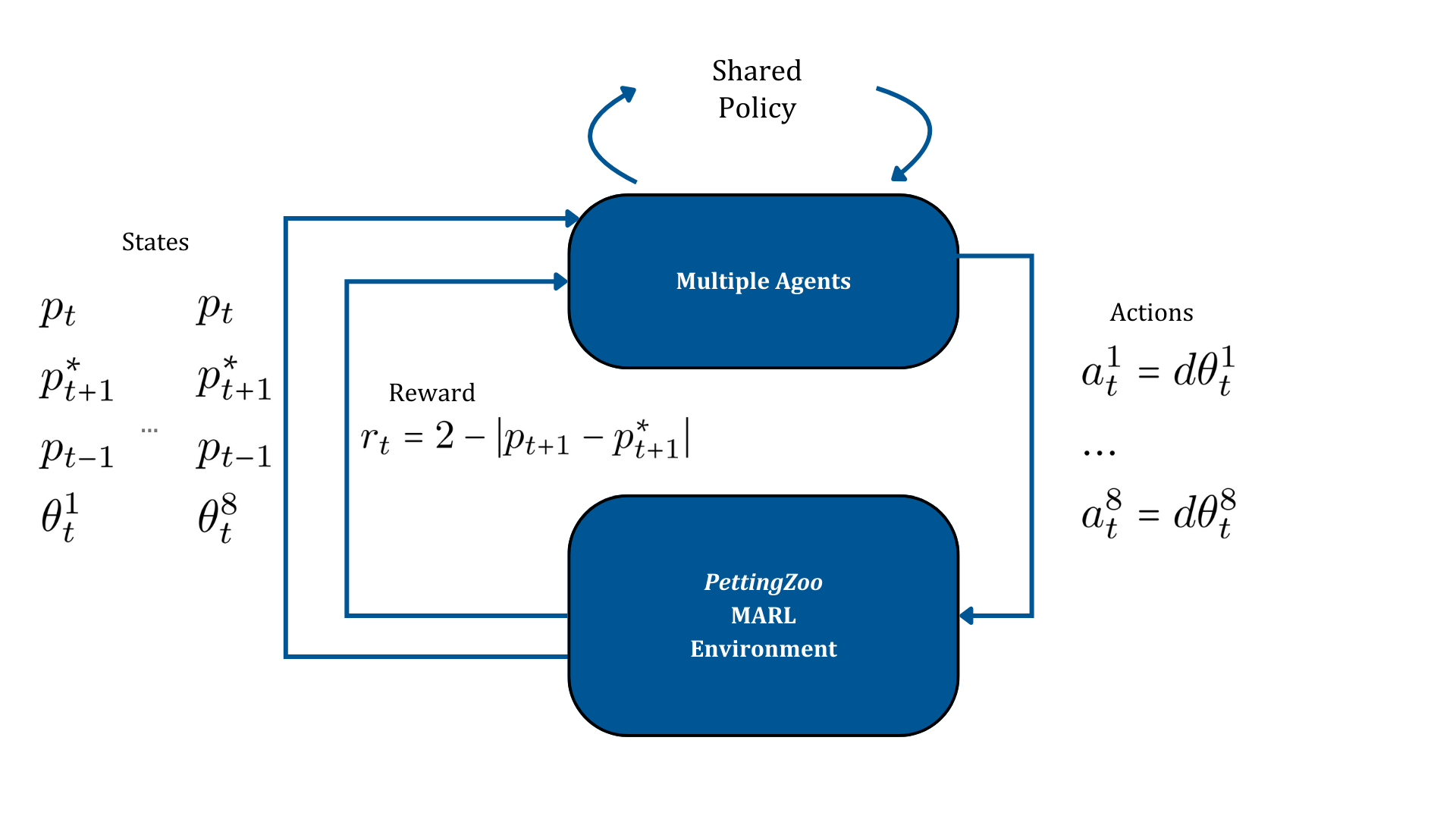}\\[-0.2cm]
\caption{MARL training loop, where $p^*$ is the target reactor power, $p$ is the measured/simulated reactor power, and $d\theta^n$ is the control drum rotation speed of the $n$th drum.}
\label{fig:marl_loop}
\end{figure}

A key challenge in MARL is nonstationarity: as multiple agents learn and update their policies in parallel, the environment dynamics continuously shift from each agent’s perspective. This makes policy optimization significantly more difficult than in single-agent RL, where the environment remains more stable. One common approach to mitigate this is single-policy training, where all agents share a common policy. This approach reduces instability during learning, since all the agents update based on shared experience rather than adapting to an evolving set of external, independently learned policies. In the context of drum-controlled microreactors, where design symmetry suggests that each control drum should behave identically under equivalent conditions, using a single policy naturally aligns with the physical constraints. A symmetry-based MARL approach has been used to train a simulated humanoid figure with bilateral symmetry to walk \cite{yan2024geometry}. Fig.~\ref{fig:marl_loop} shows a MARL training loop specific to this paper.
\hfill

Various frameworks provide standardized interfaces for MARL environments. \textit{PettingZoo} \cite{terry2021pettingzoo}, an open-source library maintained by the same organization as \textit{Gymnasium}, supports both turn-based and simultaneous-action multi-agent environments, and was used in the present work. Whereas \textit{Gymnasium} environments are defined by a state observation space, action space, and reward function, \textit{PettingZoo} environments are defined by the different agents that will interact with them, and by what kind of observation, action, and reward it will expose to each individual agent. This allows for defining very different observations and actions from one agent to another, exposing and affecting different portions of the overall environment state.

Although \textit{Stable-Baselines3} was not explicitly designed for MARL, it remains a practical choice for training single-policy MARL systems, thanks to its compatibility with \textit{PettingZoo}’s \textit{Gymnasium}-like interface. By default, a \textit{PettingZoo} environment will provide observations and collect actions from agents in a turn-based manner, computing its updated state after each agent's action. However, observations and actions may be provided to, and applied from, all agents in parallel for situations in which the environment need only update its state after actions have been collected from all the agents. When actions can be applied in parallel to create a single state update, the separate \textit{SuperSuit} \cite{SuperSuit} package provides a wrapper that appears to an RL agent in training as a vectorized environment, in which several environments are running their own simulations in parallel, as described in the previous subsection. In reality, a single environment runs a single simulation and provides observations intended for separate agents. In this way, existing single-agent RL workflows can be used to train agents in a MARL setting.

\section{Methodology}
\label{sec:methodology}

This section details the methodology used to develop and evaluate RL, MARL, and PID controllers for the Holos-Quad microreactor. All RL agents were trained in vectorized environments by using $N_{envs} = 10$, a value chosen to minimize the overall training time. Otherwise, the default hyperparameters from \textit{Stable-Baselines3} were used.

\subsection{Control Environment Interface and Performance Metrics}
\label{sec:interfacemetrics}
Having described the reactor model in Section \ref{sec:reactormodel}, the simulation of power under drum control across both short and long timescales is straightforward. Eqs.~\eqref{eq:pke}--\eqref{eq:rhoX} form a system of ordinary differential equations that may be solved given initial conditions at steady-state full power. These conditions are provided in, or are derivable from, Table \ref{tab:Holosparameter}. We chose to employ the \textit{solve\_ivp} method from \textit{SciPy}, using ``RK45,'' which is an explicit Runge-Kutta method of order 5 \cite{scipy}.

Less straightforward is interfacing this reactor model with our various controllers of interest for training and testing. Thus, we first defined a standard control environment setup through the \textit{Gymnasium} interface, which was described in Section \ref{sec:rlbackground}. For the sake of convenience, we define the standard power unit (SPU) to represent 1\% of the full operating power. \textit{For this 22 MW$_{th}$ reactor design, each SPU is equivalent to 220 kW$_{th}$.}

Observations of the current environment state $s_t$ for any controller were defined to consist of the following:
\begin{itemize}
    \item $p_t$, the current measured power (in SPU)
    \item $p^*_{t+1}$, the next desired power (in SPU)
    \item $p_{t-1}$, the previous measured power (in SPU)
    \item $\theta^1_t, ...,\theta^8_t$, the relevant current control drum positions (in angular degrees).
\end{itemize}

PID control requires a measurement and a setpoint, meaning that only the first two components of the observation are relevant for it. The last two components were added for the benefit of the RL controllers we use, which are memoryless. The previous measured power is provided as a type of observed memory. The drum position represents important control context because it determines the differential reactivity worth of each control drum in accordance with Eq.~\eqref{eq:diffdrumworth}. The term ``measured'' is noteworthy since, for the noise studies described later, the measured power can differ from the simulated power.

The only avenue for load-following control in our model is through the control drums, so actions are defined as drum speeds, given in degrees per second:

\begin{equation}
    a_t = d\theta^1_t, ...,d\theta^8_t,
\end{equation}
where $a_t$ is the action taken at time $t$. Actions are assumed to be taken at 1-second intervals, so all other quantities are also evaluated in 1-second timesteps. We constrain actuation of each of the eight control drums to less than 0.5 degrees per second, as higher speeds are unnecessary for the load-following scenarios we are examining, and since quick reactivity insertions can be a safety hazard. High-speed control actions are more likely from the emergency control rods, which are not considered in this study.

To encourage good actions and punish poor ones during the RL training process, we define the following reward function:

\begin{equation}
    r_{t} = 2 - |p_{t+1} - p_{t+1}^*|,
    \label{eq:reward}
\end{equation}
where $r_{t}$ is the reward to the RL control agent for taking action $a_t$ in state $s_t$ in order to result in state $s_{t+1}$, and $p_{t+1}$ is the measured power in that new state. The constant reward of 2 provided at each timestep encourages the agent to learn how to make it to the next timestep so as to receive another constant reward. The punishment of the resulting error in power $|p_{t+1} - p_{t+1}^*|$ can only be avoided if the RL controller learns to closely match the desired load.

During training (but not testing), episodes are terminated if the error in power exceeds 5 SPU, or if the control drums are sent to their limits of 0 and 180 degrees. In all cases, episodes are terminated if the power exceeds 110 SPU. The 5 SPU constraint during training prevents the agent from gaining experience in scenarios for which it is performing very poorly. This means that in order to make it through an entire episode to gain every timestep's constant reward of 2, the agent must at least coarsely follow the load-time profile. Minimization of the punishment then more finely tunes the agent's learning.

All training occurs based on the same 200-seconds-long power profile, or ``train,'' which ramps down from 100 to 55 SPU, then back up to 80 SPU. Other profiles are defined to help evaluate the trained and tuned controllers, all starting at steady state at 100 SPU. The ``test'' profile is also 200 seconds and features more extreme ramping. The 200-second ``low-power'' profile ramps all the way down to 30 SPU so as to test performance with greatly varying differential drum reactivities than were seen during training. The 20,000-second ``long-test'' profile contains a combination of very slow and very sharp ramps that tests how the controllers respond on much longer timescales, in which the xenon feedback in Eq.\eqref{eq:rhoX} becomes important.

To compare the varied controllers in these different settings, we define several metrics. Performance is measured based on the mean absolute error (MAE), cumulative absolute error (CAE), and control effort, as shown in the following formulas:

\begin{equation} 
\text{MAE} = \frac{1}{T} \sum_{t=1}^{T} |p_t - p^*_t|,
\label{eq:mae}
\end{equation}

\begin{equation}
\text{CAE} = \sum_{t=1}^{T} |p_t - p^*_t|,
\label{eq:cae}
\end{equation}

\begin{equation}
\text{Control Effort} = \int_0^T |u(t)| \, dt,
\label{eq:ce}
\end{equation}
where $T$ is the number of control timesteps in the episode and $u(t)$ is the control action over all relevant drums in $\degree / s$. These metrics enable direct comparison of the various RL and PID controllers when applied to the same control environment.

\subsection{Single-Action Control}
\label{sec:methodsingle}
Although all drums are handled separately in our reactor model, to compare RL to a traditional PID controller that can only handle one output signal, we limit the action space to a single drum speed that is then applied symmetrically to all eight drums. Thus, instead of $d\theta^1_t, ...,d\theta^8_t$, $a_t$ is here described with just $d\theta_t$. Accordingly, the observation space can be limited from $\theta^1_t, ...,\theta^8_t$ to just $\theta_t$, since with the single, symmetrically applied action, the drums will always be rotated identically.

The PID benchmark controller was first implemented to take actions in accordance with the following formula:

\begin{equation}
    u(t) = K_p e(t) + K_i \int_0^t e(\tau) d\tau + K_d \frac{de(t)}{dt},
\end{equation}
where $u$ is the control action, $e$ is the power error ($|p_t - p^*_t|$), and $K_p$, $K_i$, and $K_d$ are the proportional, integral, and derivative gains, respectively. These were tuned by selecting gains that minimized the CAE on the ``train'' profile. Two different algorithms from \textit{SciPy}'s \textit{optimize} module---sequential least squares programming and differential evolution---found the best gains to be $K_p = .078$, $K_i = 0$, and $K_d = .3$.

A PPO agent was trained using an $N_{timesteps}$ of 2 million on the ``train'' power profile. Because this RL controller was trained in a single-action version of the environment, we term it ``single-RL.''

Both the PID controller and single-RL agent were tested on the ``test,'' ``low-power,'' and ``long-test'' load-following profiles.

\subsection{Multi-Action RL and MARL}
\label{sec:methodmarl}
Unlike PID, RL can handle multiple input and output signals. While our use of a point kinetics reactor model prevents us from exploring the interesting control scenarios that might arise from reactor asymmetries, we make the assumption that in the absence of such asymmetries, control drum movements should be symmetrically applied. For this multi-output control scenario, we may either employ a controller that uses the global reactor state in order to decide on all eight drum movements, or eight controllers that each make local decisions about their respective drums. The first strategy is well set up for traditional RL training, while the second one describes a multi-agent setting that lends itself to treatment with MARL.

We explored and compared these two strategies by using three separate trained agents: a straightforward application of a PPO agent to provide training in the full, eight-action environment described in Section \ref{sec:interfacemetrics} (i.e., ``multi-RL''), a similarly trained ``symmetric-RL'' agent that is given a modified reward to encourage symmetric drum movements, and an agent trained in a MARL environment (i.e., ``trained-MARL''). All were trained using an $N_{timesteps}$ of 5 million based on the same 200-second ``train'' profile used to train and tune the single-RL and PID controllers.

While multi-RL has no incentive to take symmetric actions, symmetric-RL is given a penalty for asymmetric actions. This is due to the following modified reward function:

\begin{equation}
    r_{t} = 2 - |p_{t+1} - p_{t+1}^*| - k\times range(a_t),
    \label{eq:reward}
\end{equation}
where $k$ is a tunable constant. The agent must learn symmetric outputs to minimize the punishment from this final term. Ultimately, we chose to use $k=1$ since the action magnitudes are between -0.5 and 0.5 degrees per second, for a maximum penalty of 1, while the maximum penalty from the error term is 5 SPU, due to the episode termination criteria we defined. Giving higher weight to the range term incentivizes the agent to learn to take actions of 0 degrees per second, which is indeed symmetric but not the behavior we are looking for. Giving less weight results in error minimization being the only focus and defeats the point of modifying the reward to encourage symmetry.

To train the MARL agent, we developed a \textit{PettingZoo} environment based around the multi-action \textit{Gymnasium} environment. Eight agents are defined, with their action and observation spaces being limited to a single drum, while the overall reward function is kept as $r_{t} = 2 - |p_{t+1} - p_{t+1}^*|$. For this application, the spaces of potential observations and actions are identical between drums, since they only differ in terms of which drum is represented and actuated. Actions are assumed to be taken by each drum simultaneously, thus allowing for mock parallel training with \textit{SuperSuit} and \textit{Stable-Baselines3}, as described in Section \ref{sec:marlbackground}. A potential point of confusion arises from the PPO agent believing it is receiving observations from eight independent environments, due to the \textit{SuperSuit} wrapper, when in reality there is only one MARL environment. To correspond with 5 million true simulation timesteps, the agent is instructed to learn using an $N_{timesteps}$ of 40 million.

Note that the action and observation spaces of the single-RL agent from Section \ref{sec:methodsingle} and the trained-MARL agent are identical during training and deployment, using a single value for drum actions and observations, $d\theta_t$ and $\theta_t$, respectively. However, in the single-RL case, this was a globally observed $\theta_t$ with a resulting $d\theta_t$ that was then applied to all drums, whereas in the trained-MARL case, each agent receives a local $\theta_t$ specific to the drum it controls and takes a similarly local $d\theta_t$ action.

To re-summarize the differences between these agents, multi-RL and symmetric-RL are each single agents that control all eight drums at once. Meanwhile, within our MARL framework, eight separate agents train a shared policy and independently control their respective local drums.

In addition to the different evaluation load-following profiles described above, to test these methods even further, we developed scenarios in which drums were randomly disabled such that each agent had to compensate by moving the remaining drums to a greater extent in order to achieve the same result. We also tracked the progression of drum positions so as to observe their symmetry or asymmetry. Due to potentially asymmetric actions from the multi-RL controller, the ``long-test'' profile served not only to check how the controllers react to xenon feedback, but also to the long-term accumulation of uneven drum movements.

\subsection{Control with Noisy Observations}
\label{sec:methodnoise}
To evaluate the resilience of these methods, we also studied the impact of noisy observations. As mentioned in Section \ref{sec:interfacemetrics}, this required separately tracking both the true power for the purpose of the simulation and the measured power observed by the control agent. We assumed the noise to be Gaussian.

We evaluated each agent in test environments featuring increasing levels of noise, with a standard deviation of 0 to 5 SPU. Since we modeled our noise randomly, each noise level required several repetitions in order to get a true sense of performance. Thus, we chose to collect reward statistics after repeating each noise level 50 times. Five SPU was chosen as a conservatively high amount of noise, as it corresponds to over 1 MW$_{th}$ of measurement error, which is quite unlikely.

\subsection{Software and Computing Details}

The following packages (with version numbers and their dependencies) were used with Python 3.11: \textit{PettingZoo} 1.24.3, \textit{SciPy} 1.15.2, \textit{Stable-Baselines3} 2.5.0, and \textit{SuperSuit} 3.9.3. Training was performed on an Intel Xeon Platinum 8480+ CPU in Idaho National Laboratory's publicly facing WindRiver computing cluster, since the \textit{Stable-Baselines3} implementation of PPO is intended to run CPU-only, unless image processing is involved. The code for implementing this methodology in full is freely accessible on Github (see the Data Availability section at the end of this article).

\section{Results and Discussion}
\label{sec:results}
This section details and discusses the evaluation results for the RL, MARL, and PID controllers applied to various load-following scenarios. Table \ref{tab:control_metrics} summarizes the results for all agents and scenarios. Subsection 5.4 concludes this section by discussing the limitations of this work and providing a roadmap for future work in this area.

\begin{table}[!h]
    \centering
    \caption{Performance metrics for each controller on different load-following profiles.}
    \begin{tabular}{lccc}
        \hline
        & MAE (SPU) & CAE (SPU-seconds) & Control Effort (°) \\
        \hline
        PID test & 0.37 & 74.18 & 91.85 \\
        Single-RL test & 0.15 & 29.51 & 92.11 \\
        Multi-RL test & 0.13 & 26.29 & 228.30 \\
        Symmetric-RL test & 0.15 & 29.46 & 101.19 \\
        MARL test & 0.39 & 77.20 & 94.69 \\
        \hline
        PID low-power & 0.60 & 119.12 & 94.19 \\
        Single-RL low-power & 0.23 & 45.84 & 96.15 \\
        Multi-RL test & 0.21 & 41.02 & 245.59 \\
        Symmetric-RL test & 0.24 & 48.36 & 100.32 \\
        MARL test & 0.46 & 91.77 & 98.02 \\
        \hline
        PID long-test & 0.0064 & 128.09 & 96.95 \\
        Single-RL long-test & 0.26 & 5237.40 & 96.73 \\
        Multi-RL long-test & 2.22 & 44382.92 & 2470.69 \\
        Symmetric-RL long-test & 0.68 & 1057.52 & 352.37 \\
        MARL long-test & 0.19 & 3786.70 & 96.36 \\
        \hline
    \end{tabular}
    \label{tab:control_metrics}
\end{table}

\subsection{Single-Action Control}
\label{sec:resultsingle}
Although training for the single-RL agent was carried out for 2 million timesteps, the best set of parameters were trained using 1.2 million timesteps, after which the training deteriorated because the agent was unable to improve the reward signal any further. All single-RL tests reported here represent the best agent found, and the PID controller was tested under equivalent conditions.

\begin{figure}[!h]
\centering
\includegraphics[width=0.85\textwidth]{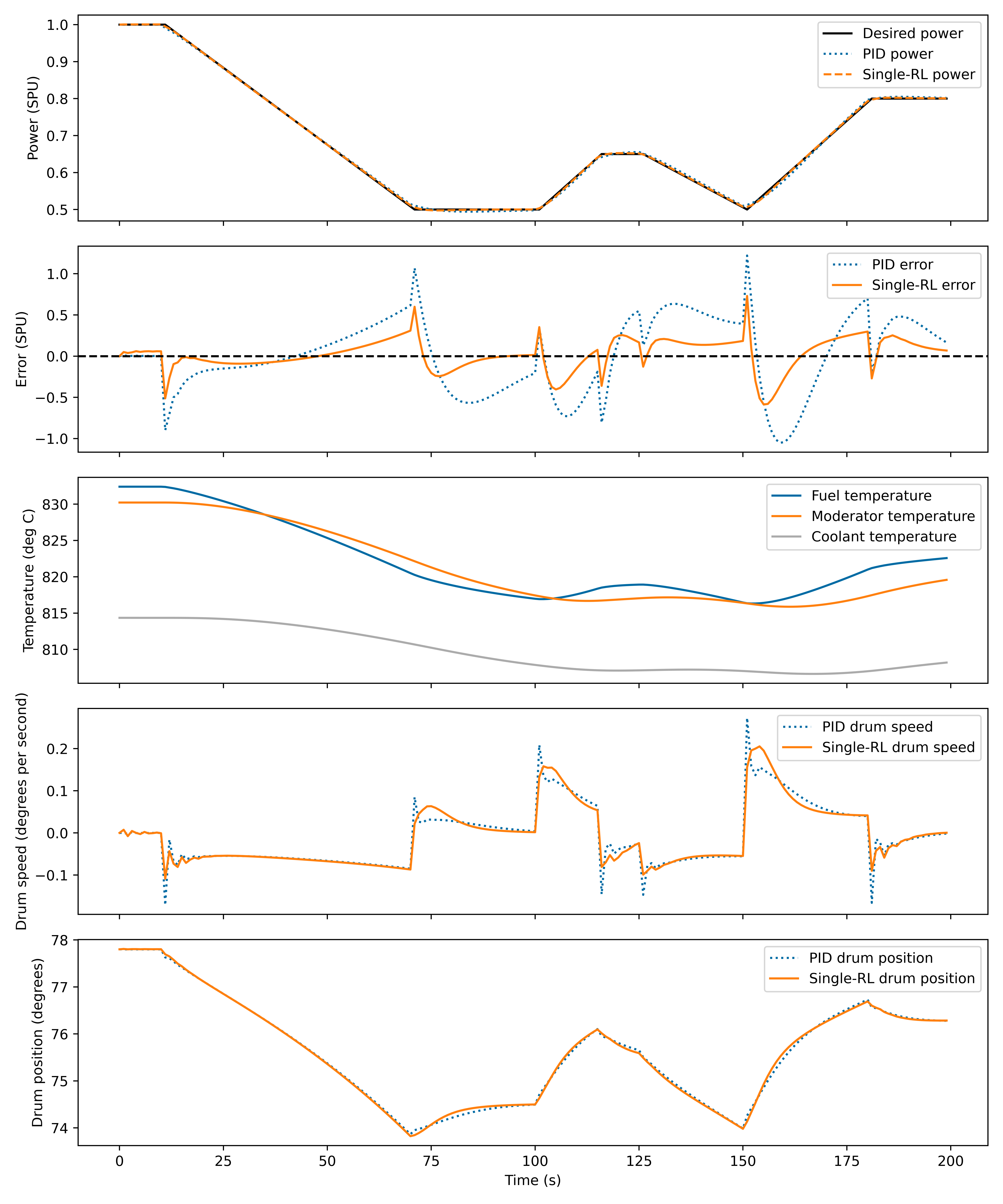}\\[-0.2cm]
\caption{Plots showing the power, error, temperature, control drum speed, and control drum angle over time, with the PID and single-RL controllers applied to the ``test'' profile. The training profile shows comparable performance and metrics, indicating no overfitting. Single-RL was trained to take a single action that is applied symmetrically to all drums.}
\label{fig:singletest}
\end{figure}

Figure \ref{fig:singletest} shows plots of several quantities of interest over a test episode under PID and single-RL control. The uppermost plot, for power over time, shows both controllers being generally able to follow the desired load. At points where ramping suddenly changes, single-RL matches the desired load more accurately, with PID unable to replicate the sharp turns in load. While these sharp changes in ramping might be unrealistic for real-world load-following situations, which would be expected to increase and decrease more smoothly, it still points to single-RL as being able to achieve significantly lower error values than PID, with comparable control effort, in scenarios that require the precision, as recorded in Table \ref{tab:control_metrics}. 

The temperature plot in the same figure shows how fuel, moderator, and coolant temperatures lag slightly behind the power and add their own feedback contributions to reactivity---contributions that the drums must then counteract. This is evidenced by continual drum movement, even during power ramp plateaus in the middle and at end of the episode. The fact that both controllers exhibit oscillatory drum speeds around points of inflection in the power ramp rate is attributable to the discrete 1 second timesteps in which actions are chosen and applied. Nevertheless, the drums are still moving slowly, within the industry constraints of lower than 1 degree/s in speed \cite{marvel2023limit}, while all the temperature values remain within the safety limits and do not experience any oscillatory changes.

\begin{figure}[!h]
\centering
\includegraphics[width=0.9\textwidth]{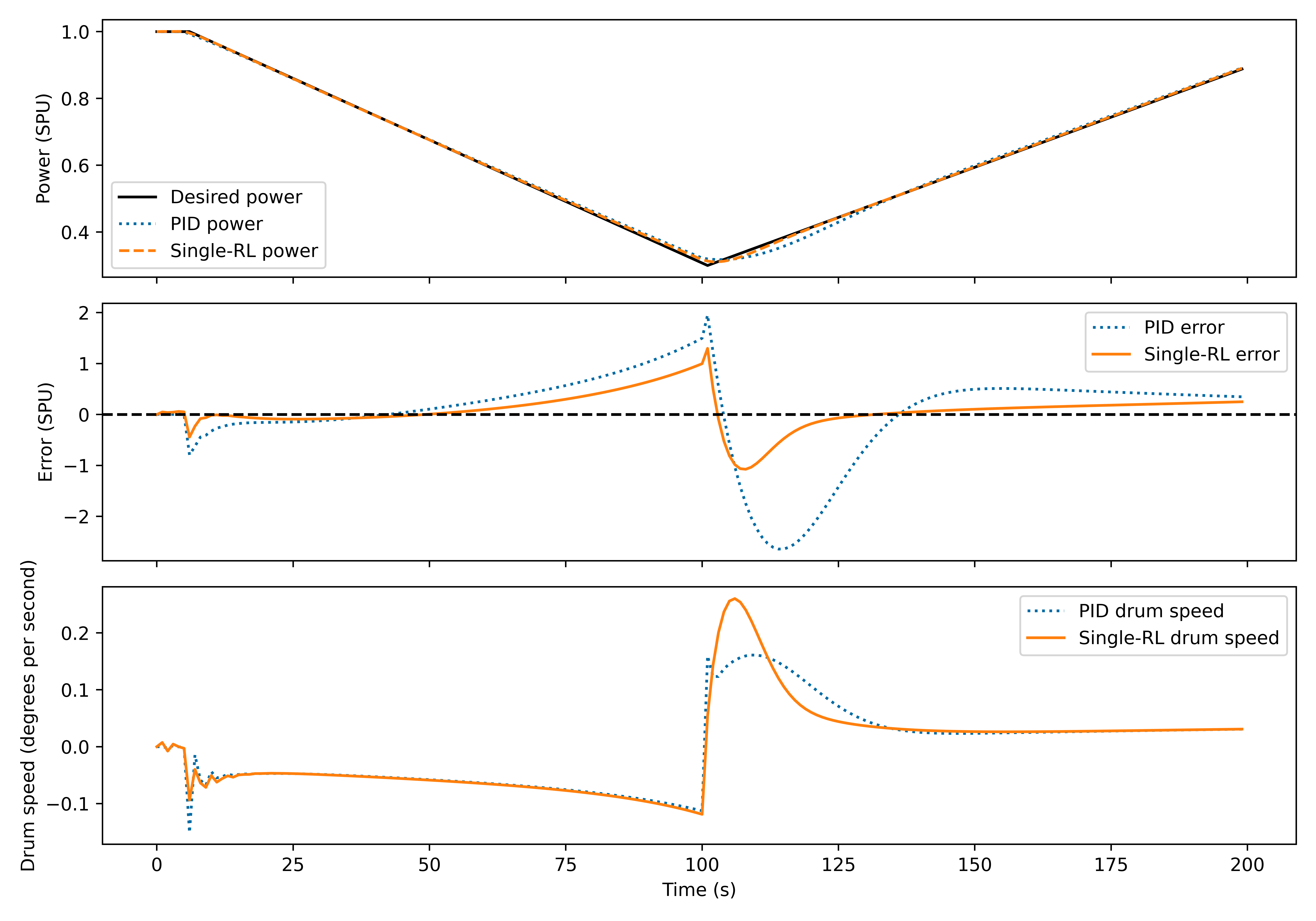}\\[-0.2cm]
\caption{Plots showing the power, error, and drum speed over time, with the PID and single-RL controllers applied to the ``low-power'' profile.}
\label{fig:singlelowpower}
\end{figure}

Single-RL once again vastly outperforms PID in the low-power profile, achieving less than half the error value of the latter, and with slightly less control effort (see Table \ref{tab:control_metrics}). This is also seen in Fig.~\ref{fig:singlelowpower}, where PID is unable to handle the sharp change in the ramp slope at 35 SPU, reaching error values of over 2 SPU. Single-RL in this scenario is even more advantageous than in the previous test profile, indicating that RL is better able to handle power levels unseen during training or tuning. This could be due to the additional drum position information that the RL agent has access to and which a PID controller is unequipped to consider.

\begin{figure}[!h]
\centering
\includegraphics[width=0.9\textwidth]{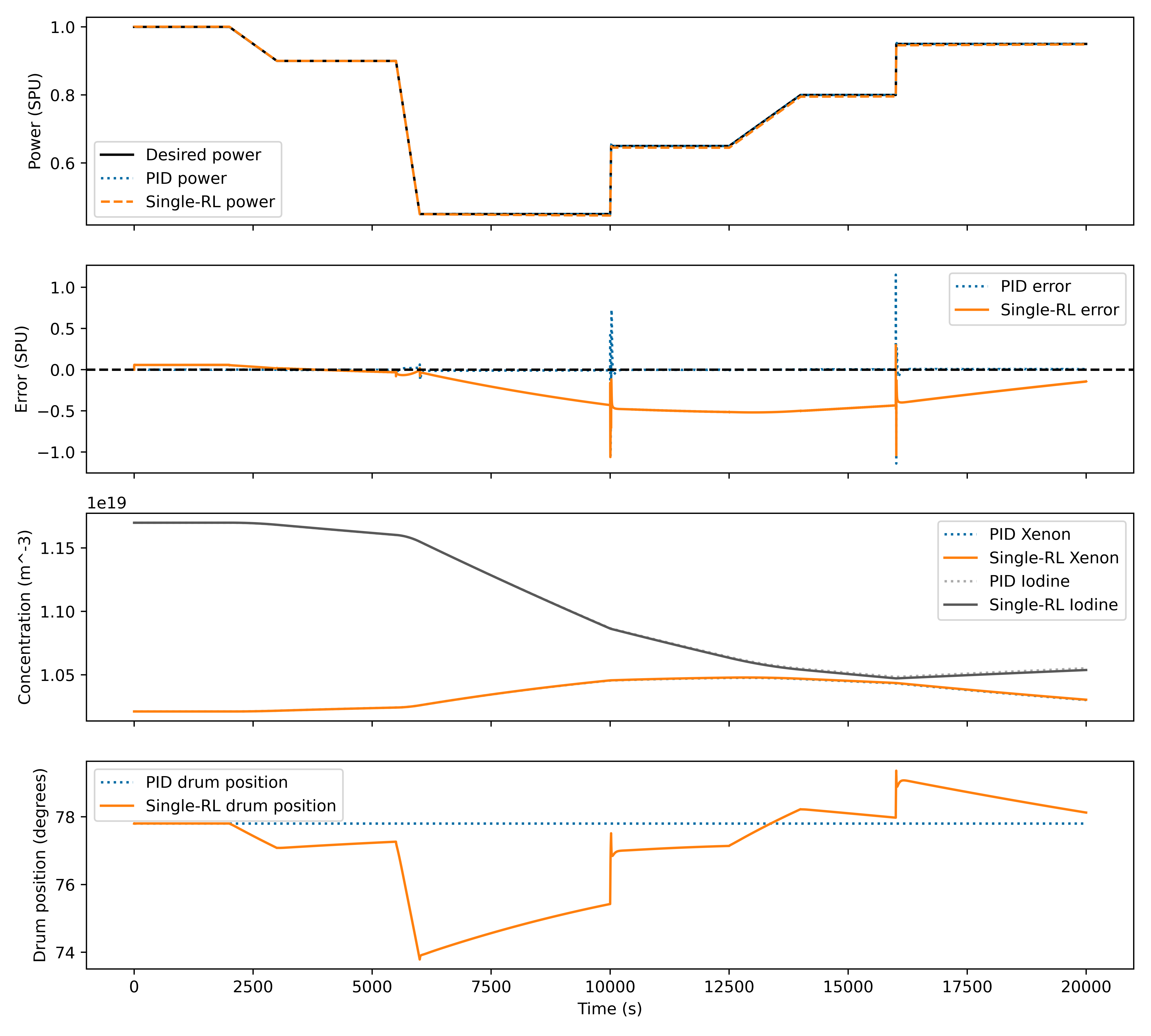}\\[-0.2cm]
\caption{Plots showing the power, error, xenon and iodine concentrations, and control drum position over time, with the PID and single-RL controllers applied to the ``long-test'' profile.}
\label{fig:singlelongtest}
\end{figure}

On the ``long-test'' profile, the results favor the PID controller, which eliminates any error signal by design. In contrast, while single-RL correctly acts to partially counteract xenon reactivity insertions during long plateaus by initiating opposing drum movements, consistent error builds up during this process, as seen in Fig.~\ref{fig:singlelongtest}. Compared to the low-power results, which were much better for single-RL than for PID, this long transient showed single-RL to be the more lackluster of the two. ML methods should generally not be expected to work outside their training parameters, so this result is none too surprising; however, it indicates that single-RL's policy does not involve directly minimizing the error signal, thus bringing up explainability concerns that will need to be addressed in future work. Nonetheless, the highest error during this ``long-test profile'' was still observed for the PID controller, whereas the RL agent consistently maintained its individual error values within 1\% (0.26\%, as given in Table \ref{tab:control_metrics}), a level deemed acceptable for this application. 

Per our analysis, the following ideas could potentially further enhance single-RL training performance. However, the authors chose not to fully implement them in this work. This decision was made to avoid overfitting the agent and to maintain a framework that is simple yet effective.

\begin{itemize}
    \item A more comprehensive training profile that includes shallower and steeper ramps than those currently provided
    \item A much longer training profile that thus includes xenon's impacts on reactivity
    \item A more complex reward function that further emphasizes reducing error during ramps by either punishing uncorrected errors or adding an extra reward for error rates within a certain tolerance
    \item Additional components to the state observation that serve to give the agent more memory (e.g., the current desired power or the previous drum position)
    \item A training profile that minimizes the length of the initial steady state so as to encourage faster learning of load following, rather than initial learning of inaction.
\end{itemize}

However, even without any of these improvements, the single-RL results provide an initial baseline for applying RL to microreactor drum control, showing very promising performance, and with less-intensive engineering of the reward function or the scenarios to which RL is exposed. Even with a simple reward and minimal hyperparameter tuning, single-RL outperforms PID on short transients and maintains less than 1 SPU of error when applied to more realistic extended scenarios. More importantly, it demonstrated excellent generalization and extrapolation to new transients across varying power levels and extended time scales, all without requiring retraining---an area that typically represents a significant limitation for ML algorithms, and for RL in particular. For instance, the authors could have conducted the RL training directly on the ``long-test'' profile shown in Fig.~\ref{fig:singlelongtest}. While this approach would likely result in longer training times and potentially superior long-term performance---possibly matching or surpassing that of PID---it would undermine the core argument that RL successfully generalizes to new long-term scenarios.

\subsection{Multi-Action RL and MARL}
\label{sec:resultmarl}
The capability to actuate control drums independently of one another would open up possibilities for better management of power/temperature asymmetries in the core. While we cannot model such asymmetries with point kinetics, here we judge RL's potential to learn such strategies in a symmetric core.

\begin{figure}[!h]
\centering
\includegraphics[width=0.9\textwidth]{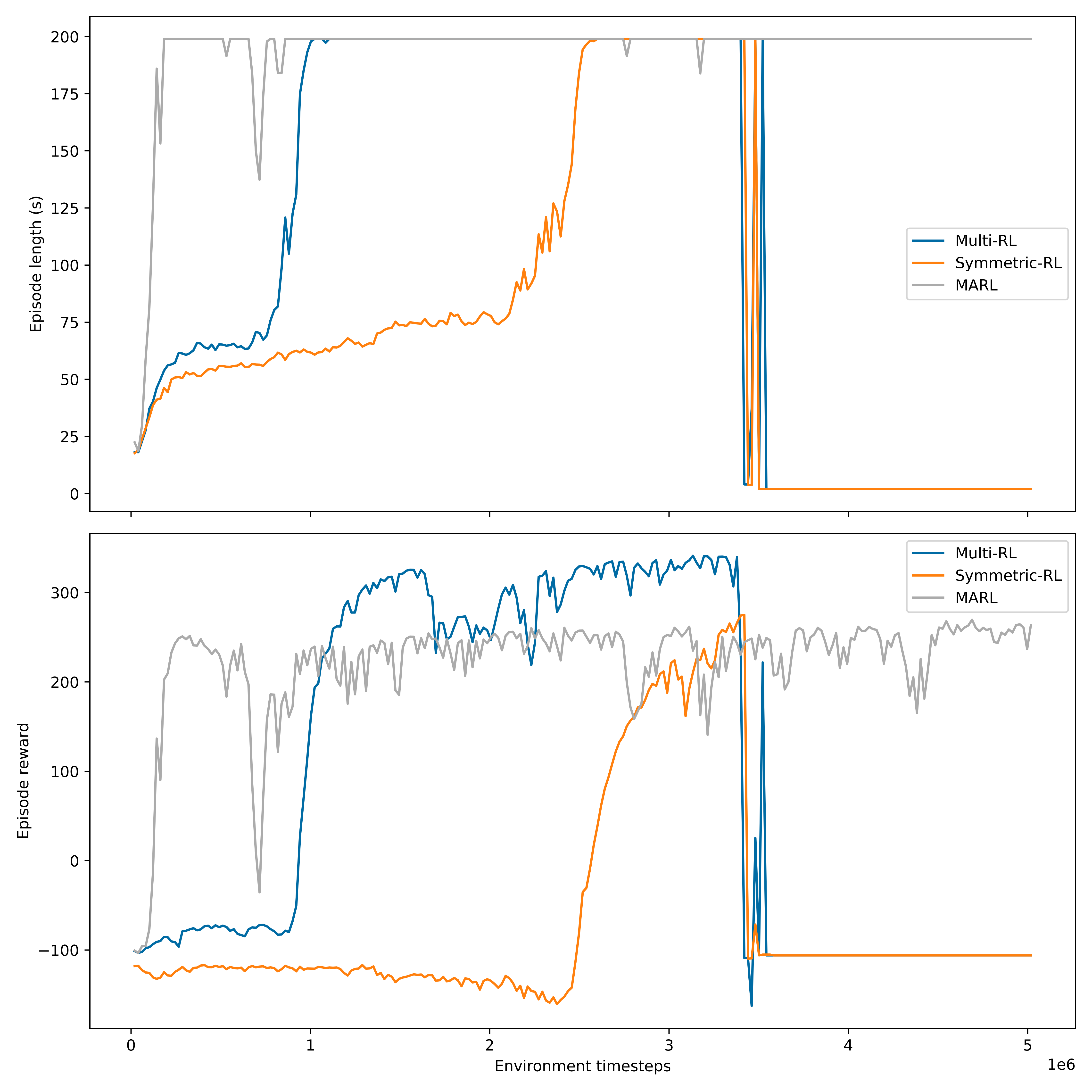}\\[-0.2cm]
\caption{Training curves for multi-RL, symmetric-RL, and MARL. Multi-RL is an agent trained to take a separate action for each of the eight control drums. Symmetric-RL, which is a type of multi-RL, is trained with a penalty for asymmetric actions. MARL is the framework to which a trained policy is applied as eight separate agents: one to independently actuate each control drum.}
\label{fig:trainingcurves}
\end{figure}

Graphs showing improved agent performance as a result of the training process can be found in Fig.~\ref{fig:trainingcurves}. While multi-RL and MARL quickly learned how to make it through the whole 200-second training episode, symmetric-RL took longer to make it past the initial steady state. In fact, it took multiple run attempts for it to train at all within the 5 million allotted timesteps, and as will be shown, it failed to learn symmetry. RL training is known to generally depend on random seeding; however, considering the use of a robust PPO agent, this is a particularly extreme case. Adjusting the reward to put less weight on the symmetry penalty leads to performance similar to that of multi-RL. Higher weighting of this penalty drastically reduces the chances that training will further progress, since to minimize the punishment stemming from uneven actions it learns to minimize its actions entirely, which is indeed symmetrical.

During \textit{Stable-Baselines3} training of PPO, the agent takes stochastic actions to assist in exploring the full policy space. Fully trained agents are deployed to take deterministic actions that exploit the trained policy. Although our MARL framework outputs symmetric actions by design during deployment, we see that during training, asymmetric actions are practically guaranteed as a result of stochasticity. This leads to a degree of nonstationarity during training, with each drum learning to act under the assumption that all the other drums will behave somewhat randomly. The result is that MARL converges to a lower maximum reward than does multi-RL in the lower half of Fig.~\ref{fig:trainingcurves}. In a real-life setting, it is arguably realistic to assume that other controllers will behave imperfectly, but the level of reward decrease is significant.

The best model for MARL was trained by using around 1.5 million timesteps---much fewer than the 2.8 million timesteps used for multi-RL and the 3.5 million for symmetric-RL. This simulation efficiency stems from the fact that one simulation timestep in the MARL framework corresponds to eight timesteps of training experience for the PPO agent, while a one-to-one correspondence holds for multi-RL and symmetric-RL. In future work involving more costly and higher-fidelity asymmetric core models, such training efficiency will be essential and will reveal the power of MARL in tackling these problems.

\begin{figure}[!h]
\centering
\includegraphics[width=0.9\textwidth]{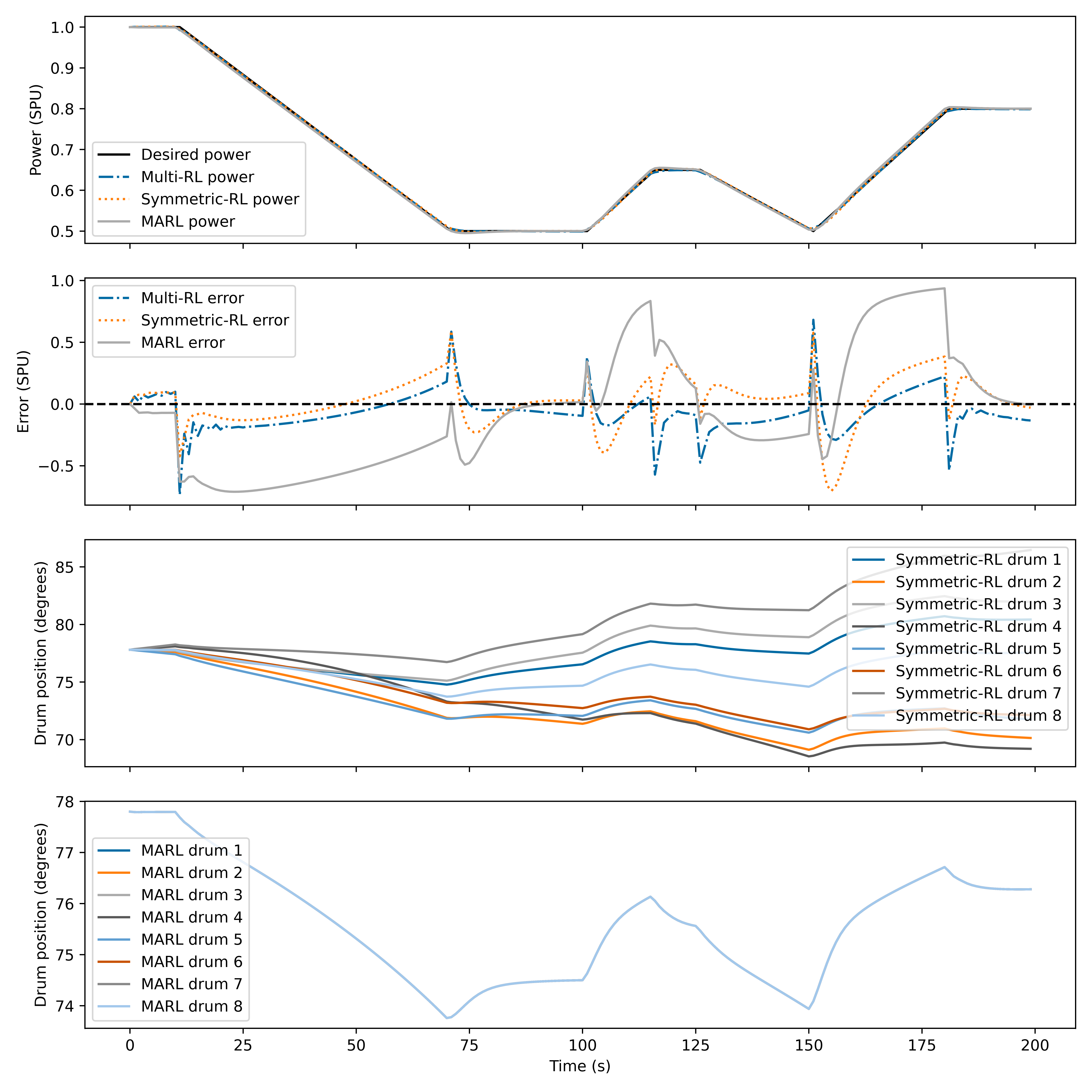}\\[-0.2cm]
\caption{Plots showing the power, error, and drum positions over time for the trained multi-RL, symmetric-RL, and MARL controllers when applied to the ``test'' profile. As the multi-RL drum positions were similar to the symmetric-RL positions, they have been omitted.}
\label{fig:multitest}
\end{figure}

Both multi-RL and MARL perform well in the test profile displayed at the top of Fig.~\ref{fig:multitest}. However, in comparing their levels of control effort (see Table \ref{tab:control_metrics}), multi-RL is clearly wasting drum movements by moving some drums in opposing directions. In reality this would create undesirable core power distributions, though this is something that cannot be modeled with point kinetics. This result is more clearly visualized in the lower plot in Fig.~\ref{fig:multitest}, showing how MARL moves all drums in perfect synchrony whereas symmetric-RL moves them in an unequal manner. Multi-RL, which had no symmetry penalty, also moves the drums unevenly. It is worth noting that while MARL shows the highest errors in this test case, when all metrics are considered, it shows similar performance to PID, while showing more physical actions that will be essential for ensuring a symmetric power distribution in the core.

\begin{figure}[!h]
\centering
\includegraphics[width=0.8\textwidth]{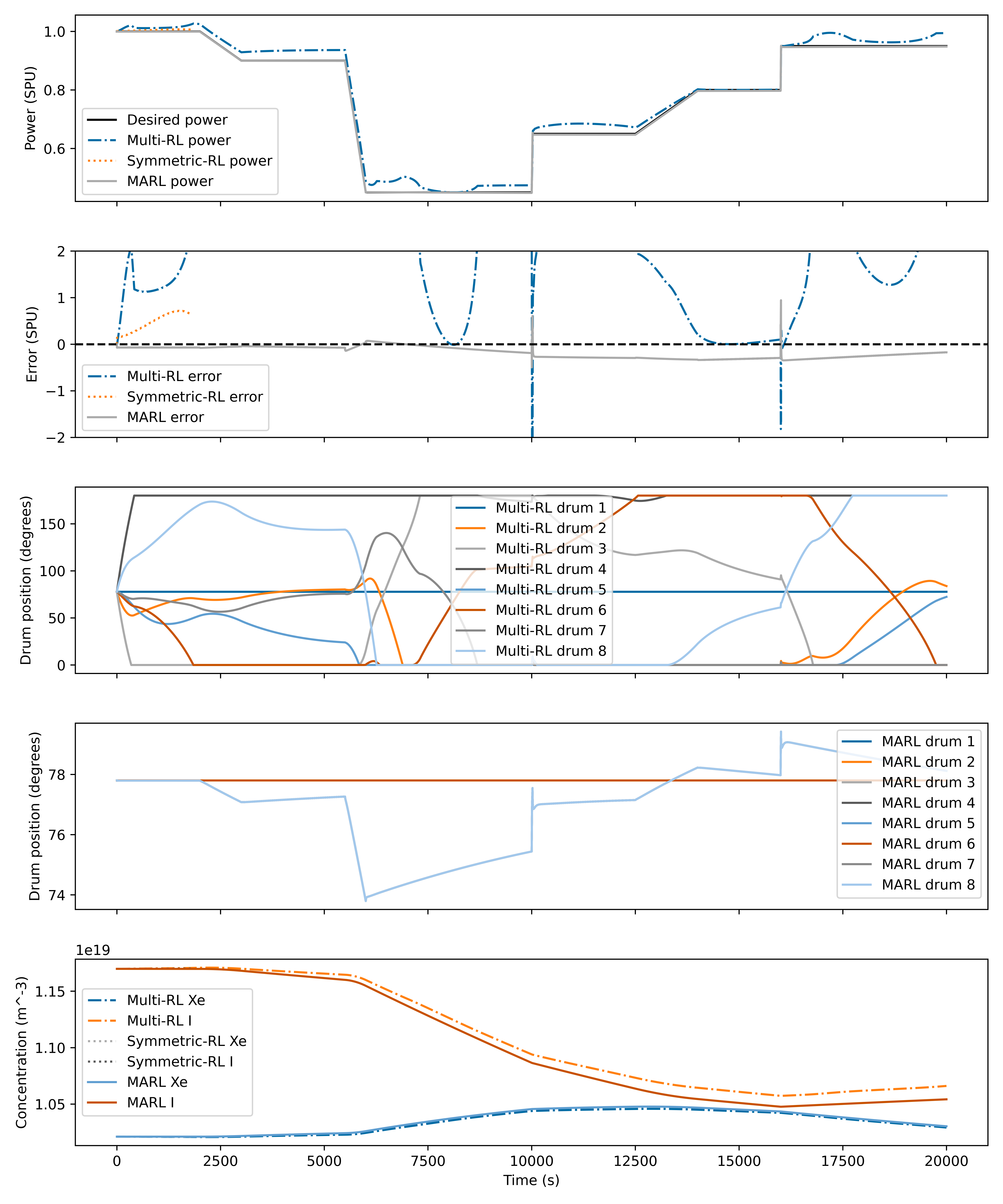}\\[-0.2cm]
\caption{Plots showing the power, error, drum positions, and xenon and iodine concentrations over time for the trained multi-RL, symmetric-RL, and MARL controllers when applied to the ``long-test'' profile with one random drum disabled from moving. Error axis bounds were selected to detail MARL performance, since the multi-RL errors are large enough to see in the power plot.}
\label{fig:multilongtest}
\end{figure}

On longer timescales, this effect is more pronounced and causes quick failure in the case of symmetric-RL, and abysmal performance in the case of multi-RL, as seen in Fig.~\ref{fig:multilongtest}. Drums in multi-RL are quickly saturated at 0 or 180 degrees (row 3 in Fig.~\ref{fig:multilongtest}), leaving few drums left for control. In this scenario, a single drum is disabled and forced to remain at its initial position throughout the episode. MARL shows, in row 4 of Fig.~\ref{fig:multilongtest}, that all the remaining drums can move in sync and make up for the disabled one.

Over long time periods, MARL shows the same accumulation of error as single-RL when xenon feedback becomes significant. However, in comparing the relevant values in Table \ref{tab:control_metrics}, the error is noticeably less for MARL. The error reduction may also stem from the unique effects of stochastic training in the MARL framework. Otherwise, similar conclusions as for single-RL can be drawn about the steps needed to improve performance on long transients for MARL, which we again avoided here so as to ensure exploration of the generalizability of MARL control from short transients in training, to testing over longer ones.

Overall, the MARL approach is the most reasonable RL method for multi-output control drum signals, as it is the only one that can reliably learn symmetric actions. It also converges much faster than the other methods, due to its efficient use of simulation experience. These two aspects will also hold true for future applications in asymmetric environments when point kinetics models are not used.

\subsection{Control with Noisy Observations}
\label{sec:resultnoise}
\begin{figure}[!h]
\centering
\includegraphics[width=0.65\textwidth]{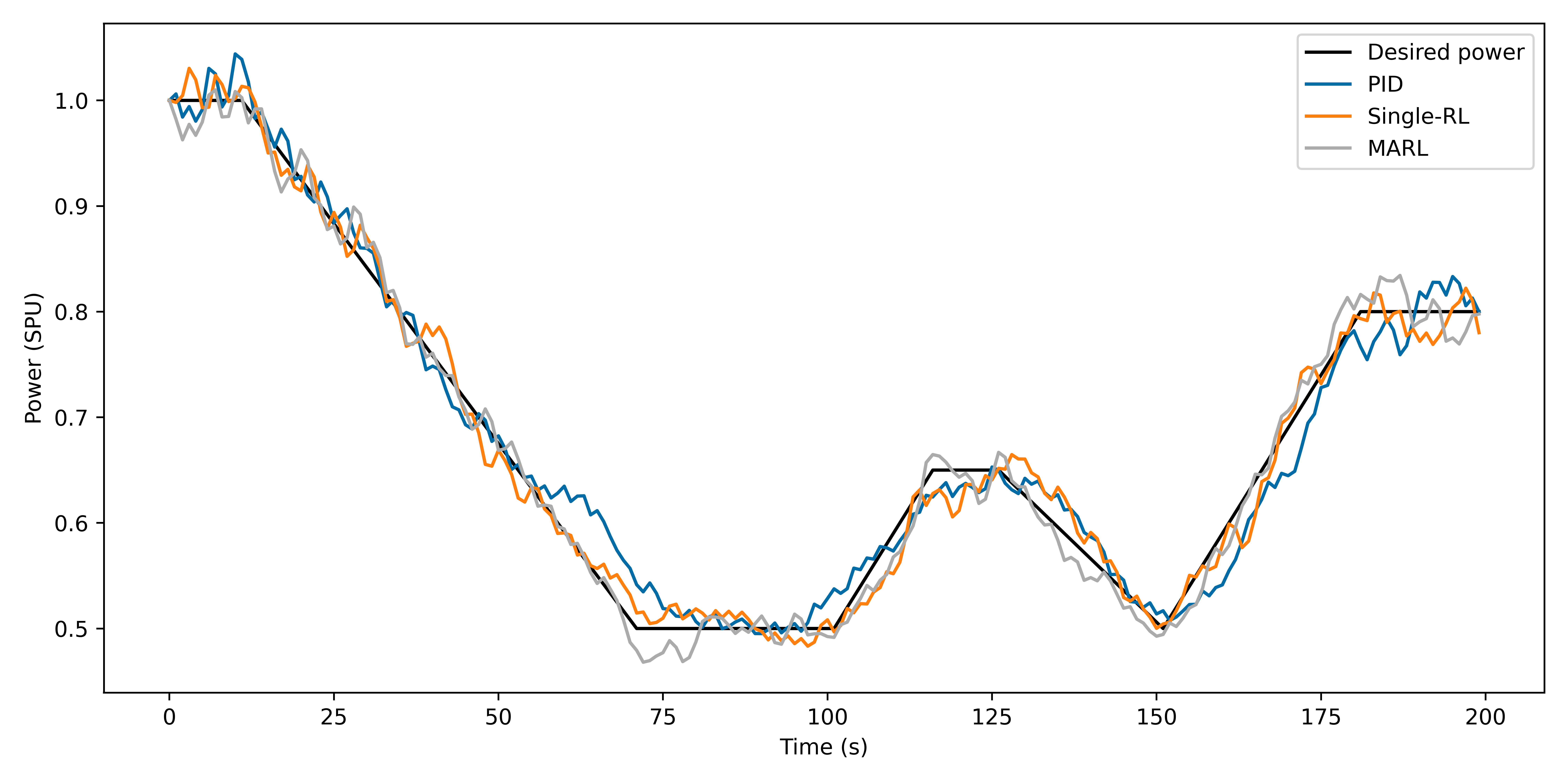}\\[-0.2cm]
\caption{Power profiles with 2-SPU-standard-deviation Gaussian noise (e.g., 440 kW$_{th}$) for the PID, single-RL, and MARL controllers on the test profile.}
\label{fig:noiseruns}
\end{figure}

For a final test of robustness, we explored adding Gaussian noise to the PID, single-RL, and MARL controllers. Fig. \ref{fig:noiseruns} shows an example of how these agents progress through episodes in which Gaussian noise with a standard deviation of 2 SPU was added to the measured power (2 SPU corresponds to around 440 kW$_{th}$).

\begin{figure}[!h]
\centering
\includegraphics[width=0.65\textwidth]{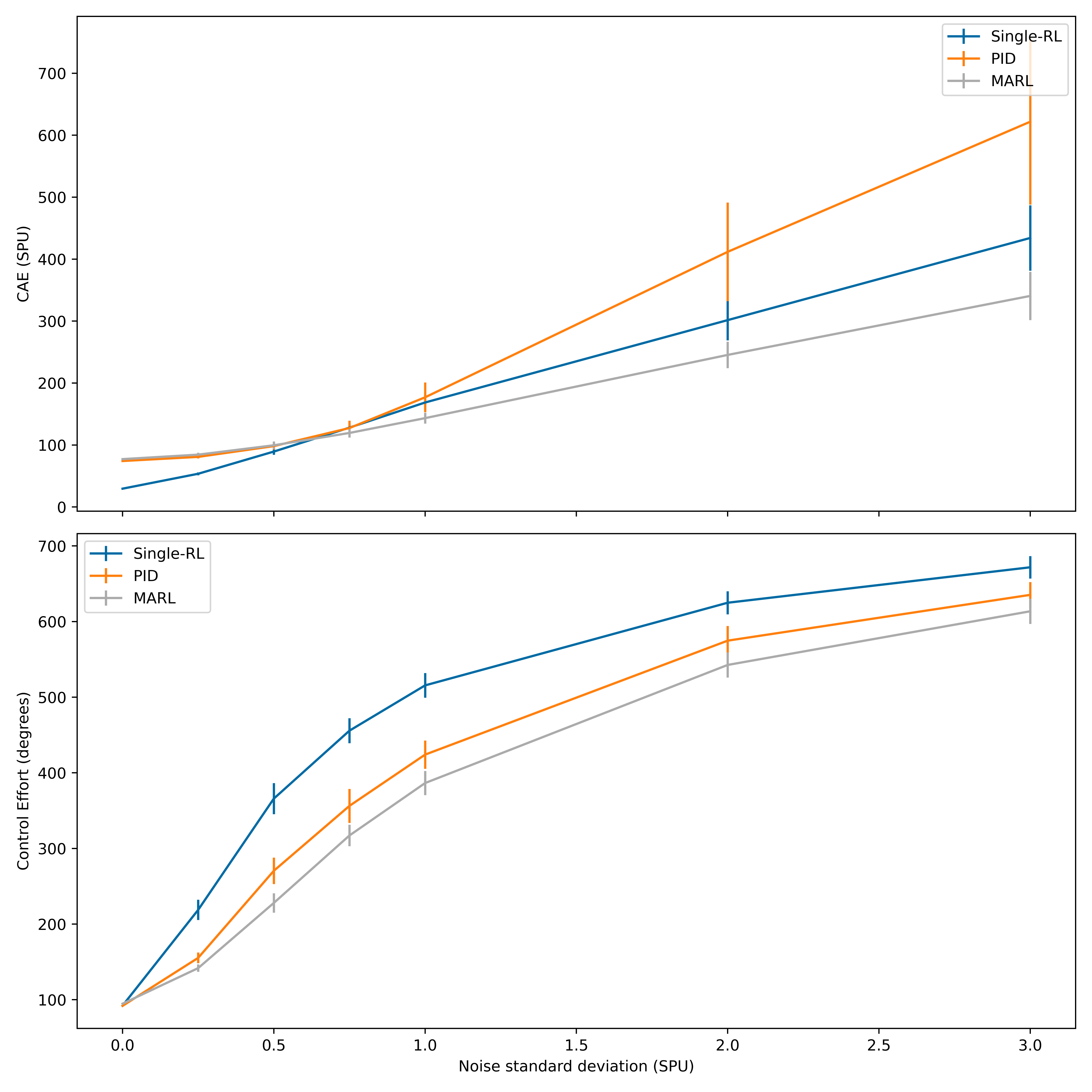}\\[-0.2cm]
\caption{Performance metrics versus increasing Gaussian noise in power measurements for the PID, single-RL, and MARL controllers. Each point represents 50 runs that have been averaged together. The bars represent the standard deviation from the 50 runs at each of these points.}
\label{fig:noisemetrics}
\end{figure}

While it seems that the PID controller performance in the middle of the profile is worse than the others, this is difficult to discern by eye from such a noisy graph, and due to randomness, performance can vary from run to run. Thus, compiling statistics from several runs is necessary to reach useful conclusions about relative controller performance. Fig.~\ref{fig:noisemetrics} plots the averages of the CAE and control effort over 50 runs, with error bars representing 1 standard deviation for several levels of Gaussian noise. It is clear that single-RL and MARL quickly begin outperforming PID as the noise is increased, implying that RL may become more robust under noisy signals or external disturbances.

For both RL-based agents, this is likely due to the extra drum position information they have access to. In MARL's case, it may also be attributable to stochastic training, in which randomized actions from other controllers acted as a sort of noise during training, thereby improving their robustness.

While these noise studies used direct applications of Gaussian noise, practical controllers are more likely to utilize filtered measurement data. Additional work exploring the use of common filtering methods or the impacts of biased measurements on power and drum angle would be an important next step. Still, the general finding that RL is more robust to noise in comparison to PID is quite promising.

\subsection{Limitations and Implications}
\label{sec:resultlimit}

This work was primarily limited by the lack of spatial modeling, due to the point kinetics model used. We were only able to investigate under the assumption that good actions should be perfectly symmetric, which is unlikely to be true over the entire lifetime of a reactor. It also, in the case of the multi-RL agent, allowed opposing drum movements to merely cancel each other out rather than create wild core power distributions. This finding is considered novel in this study, as single-agent RL can yield impressively high rewards on paper, but as we observed in Fig.~\ref{fig:multilongtest}, these results may be achieved through unphysical actions.

Though we used PID as a benchmark, it is important to recognize its limitations as a linear controller. PID is not designed to achieve optimal control in nonlinear systems such as nuclear reactors, particularly in scenarios involving strong feedback effects from temperature and xenon. More advanced model-based control approaches such as MPC would be expected to outperform PID by explicitly considering system dynamics over a given prediction horizon. However, the computational cost of MPC makes it impractical for real-time control of reactors with complex spatial dynamics, as was observed in a recent study \cite{abdulraheem2025holos}. The PID controller should thus be viewed primarily as a reference point for assessing the RL methods. The real potential of RL and MARL lies in controlling the 2-D spatial distribution of power in a reactor, a task for which model-based methods would be prohibitively slow in regard to real-time decision-making.

As the field continues to advance toward higher-fidelity multiphysics models for microreactors \cite{price2024multiphysics,price2023thermal}, these models are expected help resolve spatial resolution challenges. However, they also introduce the problem of immense computational cost, especially when considering RL, which demands a substantial amount of experience in order to be effective. Even a 2-D simulation of a core with coupled neutronics and thermal-hydraulics can be computationally prohibitive, given that RL often requires over a million timesteps before performance improves. While it is more feasible that future RL work will continue training with point kinetics, validation in high-fidelity transient scenarios will likely be necessary. Surrogate modeling approaches, such as Sparse Identification of Nonlinear Dynamics with Control (SINDYc) \cite{brunton2016sparse}, neural networks \cite{kalise2024multi,radaideh2025multistep,saleem2020application}, uncertainty-aware methods (e.g., Gaussian processes) \cite{radaideh2020surrogate}, or even classical machine learning methods through AutoML \cite{myers2025pymaise} may represent promising alternatives for reducing computational burdens. However, these methods still require extensive data generation and must be transient-capable in order to support RL training effectively.

Lack of spatial resolution also limits the conclusions we may draw from the disabled drum tests presented in Section \ref{sec:resultmarl}, as in reality we would expect that drums closer to the disabled one would need to compensate to a greater extent than would the ones further away. Studies (described in Section \ref{sec:intro}) that used nodal point kinetics models for reactor control research focused on resolving the core with nodes axially, since for traditional large, rod-controlled reactors, imbalances primarily occur in this dimension. Drum-controlled microreactor studies would instead benefit from radial and angular nodal resolution. A fuel depletion study of disabled drums could also verify the extent to which having multiple drums provides some level of redundancy, and how long the reactor could still be successfully operated afterward. This could be particularly relevant to drum-controlled reactors utilized in space, where maintenance is impractical.

Another limitation stems from the fact that, in reality, actions from independent drums would not be made perfectly simultaneously, nor in exact 1-second timesteps. This is particularly problematic for the observations we used of the measured power at the previous timestep, as well as for the power desired at the next timestep, since the agent counts on these referring to precisely 1 second in the past and future, respectively. Observations agnostic of absolute time would be more suitable for actual deployment. Further study should consider the effects of actions taken at varying times, and could leverage \textit{PettingZoo}'s default turn-based environment setup. It would also be important to see the extent to which this approach would slow down training.

This work made no attempt to investigate RL's performance in accident scenarios that might involve superprompt criticality. While this is justified based on the presence of a separate system of emergency control rods, a clear understanding of the capabilities and limits of RL in such situations will be crucial for RL to be seriously considered in nuclear reactor control.

Overall, RL methods show promise for making effective real-time decisions and being robust to noise. The MARL approach of training independent drums should be applicable to any reactor with control symmetry (which is typical for most reactor designs), and is the most relevant RL method for training multiple controllers in this setting. With a reactor model capable of modeling core asymmetries, the observation space of each drum should be relative to its location, without reference to absolute reactor positions, such that the same policy can control each one. Since independent actions in an asymmetric environment introduce greater nonstationarity, the stochastic training issue previously described will need to be mitigated or eliminated. This will likely require usage of MARL-specific training algorithms beyond the traditional RL methods of \textit{Stable-Baselines3}.

\section{Conclusions}
\label{sec:conclusion}
This study demonstrated the feasibility of using deep RL for real-time drum control in nuclear microreactors, providing a novel approach to autonomous reactivity regulation. Through systematic evaluation, we showed that RL-based controllers, including both single-agent and MARL approaches, can achieve comparable or superior load-following performance as traditional PID controllers. In particular, RL outperformed PID in short transients and in scenarios with significant measurement noise, thus highlighting its robustness and adaptability. MARL further demonstrated the ability to maintain reactor symmetry constraints while achieving efficient, decentralized control of multiple drum actuators. However, long-duration transients that incorporated xenon feedback revealed areas for improvement, suggesting that future work should explore extended training scenarios and enhanced reward functions. Nevertheless, this study demonstrated that RL can effectively generalize from training on short transients to longer ones, enabling significant reductions in training costs. It also highlighted RL's ability to extrapolate to new scenarios, indicating reduced overfitting. Additionally, RL methods exhibited greater robustness to signal noise in comparison to PID controllers.

While this work establishes a strong foundation for RL-based control of nuclear microreactors, further research is needed to address its limitations, particularly the reliance on simplified point kinetics models. Future studies should validate RL controllers within high-fidelity multiphysics simulations and, ultimately, experimental settings. An additional angle of research could involve leveraging generative models like diffusion models \cite{yang2023diffusion} or generative adversarial networks \cite{nabila2025data} to generate quick experiences and trajectories for RL training from high-fidelity simulations. Additionally, expanding MARL applications to asymmetric reactor configurations and optimizing RL policies for safety-critical scenarios will be crucial for practical deployment. These advancements will help bridge the gap between theoretical RL applications and real-world autonomous reactor control, paving the way for safer, more efficient nuclear microreactor operations.

\section*{Data Availability}
\label{sec:data_avail}

Currently, the authors possess, in a private GitHub repository, all the data and codes needed to reproduce all the results in this work. To ensure confidentiality of this research, the authors will make this repository public during an advanced stage of the review process, and it will be listed under our research group's public Github page: \url{https://github.com/aims-umich}.

\section*{Acknowledgment}
This work is supported through the INL Laboratory Directed Research \& Development (LDRD) Program (Award Number 24A1081-116FP) for the project named ``Uncertainty Quantification Approach for Digital Twin-based Autonomous Control'' under DOE Idaho Operations Office Contract DE-AC07-05ID14517. This work is also sponsored by the Department of Energy (DOE) Office of Nuclear Energy's Distinguished Early Career Program (Award Number DE-NE0009424), which is administered by the Nuclear Energy University Program (NEUP). INL HPC resources were used for this paper.

\section*{CRediT Author Statement}

\begin{itemize}

    \item \textbf{Leo Tunkle}: Conceptualization, Methodology, Software, Validation, Formal Analysis, Visualization, Investigation, Data Curation, Writing - Original Draft.
    \item \textbf{Kamal Abdulraheem}: Methodology, Data Curation, Software, Validation, Formal Analysis, Writing - Review and Edit.
    \item \textbf{Linyu Lin}: Methodology, Funding Acquisition, Project Administration, Writing - Review and Edit.
    \item \textbf{Majdi I. Radaideh}: Conceptualization, Methodology, Investigation, Funding Acquisition, Supervision, Project Administration, Writing - Review and Edit. 

\end{itemize}


\bibliographystyle{ans_js}
\setlength{\bibsep}{0pt plus 0.3ex}
{\small
\bibliography{references}}

\end{document}